# Global Inequality in Cooling from Urban Green Spaces and its Climate Change Adaptation Potential

Yuxiang Li[1], Jens-Christian Svenning[2], Weiqi Zhou[3,4,5], Kai Zhu[6], Jesse F. Abrams[7], Timothy M. Lenton[7], Shuqing N. Teng[1,*], Robert R. Dunn[8,*], Chi Xu[1,9,*]

[1]School of Life Sciences, Nanjing University, Nanjing, China.

[2]Center for Ecological Dynamics in a Novel Biosphere (ECONOVO) & Center for Biodiversity Dynamics in a Changing World (BIOCHANGE), Department of Biology, Aarhus University, Aarhus, Denmark.

[3]State Key Laboratory of Urban and Regional Ecology, Research Center for Eco-Environmental Sciences, Chinese Academy of Sciences, Beijing, China.

[4]University of Chinese Academy of Sciences, Beijing, China.

[5]Beijing Urban Ecosystem Research Station, Research Center for Eco-Environmental Sciences, Chinese Academy of Sciences, Beijing, China.

[6]Institute for Global Change Biology and School for Environment and Sustainability, University of Michigan, Ann Arbor, MI, USA.

[7]Global Systems Institute, University of Exeter, Exeter, UK.

[8]Department of Applied Ecology, North Carolina State University, Raleigh, NC, USA.

[9]Breeding Base for State Key Lab. of Land Degradation and Ecological Restoration in northwestern China; Key Lab. of Restoration and Reconstruction of Degraded Ecosystems in northwestern China of Ministry of Education, Ningxia University, Yinchuan, China.

*Authors for correspondence: Shuqing N. Teng (shuqingteng@nju.edu.cn), Robert R. Dunn (rrdunn@ncsu.edu) and Chi Xu (xuchi@nju.edu.cn)

**Heat extremes are projected to severely impact humanity and with increasing geographic disparities. Global South countries are more exposed to heat extremes and have reduced adaptation capacity. One documented source of such adaptation inequality is a lack of resources to cool down indoor temperatures. Less is known about the capacity to ameliorate outdoor heat stress. Here, we assess global inequality in green infrastructure, on which urban residents critically rely to ameliorate lethal heat stress outdoors. We use satellite-derived indicators of land surface temperature and urban green space area to quantify the daytime cooling capacity of urban green spaces in the hottest months across ~500 cities with population size over 1 million per city globally. Our results show a striking contrast with an about two-fold lower cooling capacity in Global South cities compared to the Global North (2.1±1.3 °C vs. 3.8±2.0 °C). A similar gap occurs for the cooling adaptation benefits received by an average urban resident (Global South 1.9±1.2 °C vs. North 3.6±1.8 °C), i.e., accounting for relative spatial distributions of people and urban green spaces. This cooling adaptation inequality is attributed to the discrepancies in urban green space quantity and quality between Global North and South cities, jointly shaped by natural and socioeconomic factors. Our analyses suggest vast potential for enhancing outdoor cooling adaptation while reducing its global inequality through expanding and optimizing urban green infrastructure.**

Heat extremes are projected to be substantially intensified by global warming[1, 2], imposing a major threat to human mortality and morbidity in the coming decades[3-7]. This threat is particularly concerning as most humans now live in cities[8], including ones suffering some of the hottest climate extremes. Cities face two forms of warming: warming due to climate change and warming due to the urban heat island effect[9-11]. These two forms of warming have the potential to be additive. Climate change in itself is projected to result in rising maximum temperatures above 50 °C for a considerable fraction of the world if 2 °C global warming is exceeded[2]; the urban heat island effect will cause up to >10 °C additional (surface) warming[12]. Exposures to temperatures >35 °C with high humidity or >40 °C with low humidity can lead to lethal heat stress for humans[13]. Even before such lethal temperatures are reached, the effectiveness of workers and general health and well-being can suffer[14, 15]. Heat extremes are especially risky for people living in the Global South[16, 17]. Climate models project that the lethal temperature thresholds will be exceeded with increasing frequencies and durations, and such extreme conditions are concentrated in low-latitude regions[18, 19]. These low-latitude regions overlap with the parts of the Global South where population growth rates are high, particularly in cities, such that the number of people exposed to these conditions will increase, all things equal[20, 21]. That population growth will be accompanied by expanded urbanization and intensified urban heat island effects[22, 23], potentially exacerbating future Global North-Global South heat stress exposure inequalities.

Fortunately, we know that heat stress can be buffered, in part, by vegetation in cities[24]. Urban green spaces, and especially urban forests, have proven an effective means through which to ameliorate heat stress through shading and transpirational cooling[25-27]. The buffering effect of urban green space is influenced by its area relative to the area of the city and its spatial configuration[28, 29]. However, the effect of green space is also mediated by differences among regions, whether in their background climate, composition of green spaces, and other factors[28, 30-34]. The geographic patterns of the buffering effects of green spaces, whether due to geographic patterns in their areal extent or region-specific effects, have so far been poorly characterized.

Ideally, the cities likely to be most impacted by climate change should be those where we invest the most in creating, conserving, and restoring parks and other vegetated areas, as the value of green space in such cities is the greatest. These benefits of green space have the potential to reduce global inequalities in warming effects (if poorer cities are also greener). On the other hand, a variety of mechanisms might conspire to make the area and effects of green space less in poorer cities. A well-documented mechanism is the 'luxury effect' that describes negative correlations of urban green space (and biodiversity) and socioeconomics[35-37], wherein poorer neighborhoods tend to have less green spaces. If poorer cities also have less green spaces (e.g., via an between-city luxury effect), the Global South may have the least potential to mitigate the combined effects of climate warming and urban heat islands, leading to exacerbated and rising inequalities in heat exposure[38]. Which of these phenomena prevails is an empirical question that has not yet been adequately addressed.

Here, we assess the global inequality in the cooling capability of existing urban green spaces ('urban green infrastructure'). We used remotely sensed data to quantify 1) the cooling amelioration capacity of existing urban green infrastructure ('cooling capacity') for ~500 major cities across the world. The cooling capacity of a city is the potential of a city's green space to cool its inhabitants, assuming those inhabitants are randomly distributed across the city. We further account for the spatial distributions of urban green infrastructure and populations to quantify 2) the benefit of cooling mitigation received by an average urban inhabitant ('cooling benefit') in each city given their location. Cooling benefit is a more direct measure of the cooling realized by humans, after accounting for the geography of green space and human density. We reveal striking inequalities of both cooling capacity and cooling benefit, reflected by a two-fold gap between Global South and North cities on average. Further, we show this inequality is attributed to the differences in both 'quantity' and 'quality' of existing urban green infrastructure. Importantly, we find that there are huge potentials for many cities to enhance the cooling capability of their urban green infrastructure, and that achieving these potentials would reduce global inequalities in adaptation to outdoor heat stress.

**Quantifying cooling inequality**

Our analyses showed that increases in urban green space (i.e., urban vegetation cover measured by the remotely sensed indicators of spectral greenness, NDVI or EVI, see Methods) can effectively cool down daytime land surface temperatures during peak summer consistently across the ~500 big cities globally (Extended Data Figs. 1-2), i.e., greener cities are cooler cities. However, for different cities, an increase of per unit green space does not lead to an equal temperature decrease (i.e., different 'cooling efficiency', see Methods). The extent to which existing green infrastructure can cool down an entire city's surface temperatures thus depends on both total area and the 'cooling efficiency' of its green space, where a city's cooling efficiency is a measure of the extent to which a given proportional increase in the area of green space leads to a decrease in temperature (see Fig. 1 for examples).

Our result showed that in an average city, green infrastructure ameliorates heat stress by ~2.7 °C of surface temperature on a whole-city level (see Methods, Fig. 2). Conversely, the absence of green infrastructure is associated with increases of ~2.7 °C surface temperature across a city. This cooling capacity of green infrastructure differs greatly among cities (s.d. = 1.8 °C). Cities closer to the equator - which is to say tropical and subtropical cities - tend to have relatively weak cooling capacities (Fig. 2a, b). As Global South countries are predominantly located at low latitudes, this pattern leads to a situation in which Global South cities have, on average, approximately a two-fold lower cooling capacity than cities of the Global North (2.1±1.3 vs. 3.8±2.0 °C, Wilcoxon test, $p$ <2.2e-16; Fig. 2c).

When we account for the location of green space relative to humans in cities, the 'cooling benefit' of green space realized by an average urban resident generally becomes slightly lower than is suggested by cooling capacity (see Methods; Extended Data Fig. 3). Because urban residents tend to be densest in the parts of cities with less green spaces, the average urban resident experiences less cooling amelioration than might be expected across a city. However, such within-city-scale heterogeneity has a minor effect on the global-scale inequality. Our results show that the geographic trends in cooling capacity and cooling benefit are similar: mean cooling benefit also presents a two-fold gap between Global South and North cities (1.9±1.2 vs. 3.6±1.8 °C, Wilcoxon test, $p$ <2.2e-16; Extended Data Fig. 3c). When walking outdoors, an average person in an average Global South city receives half as much cooling amelioration from green spaces as a person in an average Global North city. The high cooling amelioration capacity and benefit of the Global North cities is heavily influenced by Northern America (Canada and the US) which have both the highest cooling efficiency and the largest area of urban green space, followed by Europe (Extended Data Fig. 4).

One way to illustrate the global inequality of cooling capacity or benefit is to separately look at the cities that are most and least effective in ameliorating outdoor heat stress. Our results showed that ~80% of the 50 most effective cities (with highest cooling capacity) are located in the Global North, while ~90% of the 50 least effective are Global South cities (Fig. 3, Extended Data Figs. 5, 6). This is true without taking into account the differences in the background temperatures and climate warming of these cities, which are likely to exacerbate these effects; cities in the Global South are likely to be closer to human thermal limits such that the ineffectiveness of green space in those cities in cooling will lead to greater health effects on humans. While the Global North cities currently accommodate larger population sizes (per city; Extended Data Fig. 5), the Global South cities commonly present higher population densities (Fig. 3) and are projected to present faster population growth[39]. This situation will plausibly intensify the urban heat island effect, but on the other hand, this also means high potential of increasing cooling benefit via increasing urban green spaces, in the sense that more people can receive the cooling mitigation from a given new neighboring green space if they live closer to each other.

Inequalities were found not only between the Global North and South, but also among cities *within* the Global South. The inequalities among the Global South cities in access to cooling green space were greater than that among the Global North cities, as measured by higher among-city Gini of cooling capacity and cooling benefit (Supplementary Fig. 1). Some Global South, subtropical cities have high green space area and hence high cooling capacity. These cities, such as Medan (Indonesia) and Manila (Philippines) will be important to study in more detail, to shed light on the mechanistic details of their cooling abilities as well as the sociopolitical and other factors that facilitated their high green area coverage (Extended Data Figs. 7, 8).

**Influencing factors**

To understand what caused the global inequality of cooling adaptation, we separately examined the effects of cooling efficiency and total green space area, representing 'quality' and 'quantity' of urban green infrastructure respectively. The simplest null model is one in which cooling capacity (at the city level) and cooling benefit (at the human level) are driven primarily by the proportional area in a city dedicated to green space. Indeed, we found that both cooling capacity and cooling benefit were strongly correlated with urban green space (Fig. 4, Extended Data Fig. 9). This is useful with regards to practical interventions; in general cities that invest in saving or restoring more green space will receive more cooling benefit from that green space regardless of where they are located. Meanwhile, the area of green space in a city (again, in proportion to its size) was positively correlated with GDP and city area. It is well known that wealth and green space are positively correlated within cities[40-44]; here we show a similar effect among them. In addition, larger cities had proportionally more green spaces, an effect that may be due to the tendency of large cities (particularly in the US and Canada) to be lower density. Cities that were hotter and had more topographic variation tended to have fewer green spaces and those that were more humid tended to have more green spaces. Given that temperature and humidity are highly correlated with the geography of the Global South and Global North, it is difficult to know whether these effects are due to temperature and precipitation *per se* or are associated with historical, cultural and colonial differences that loosely track climate.

Differences among cities in cooling efficiency played a more minor role in determining the cooling capacity and benefit of cities (Fig. 4, Extended Data Fig. 9). Most variation among cities in their cooling efficiency was unexplained. However, cooling efficiency was modestly influenced by temperature, humidity, and GDP. The warmer a city, the greater the cooling efficiency in that city. Conversely, the more humid a city the less the cooling efficiency of that city. In addition, cities with a higher GDP tended to have higher cooling efficiencies. Just why this might be deserves further exploration.

Our analyses suggest that the lower cooling adaptation capabilities of Global South cities can be explained by their lower quantity of urban green space and, to a much lesser extent, the reduced cooling efficiency of that green space (Extended Data Fig. 2). These patterns appear to be in part structured by GDP, but are also associated with climatic conditions[36], and other factors. A key question, unresolved by our work, is whether the climatic correlates of the size of green space areas in cities are due to the effects of climate *per se* or if they, instead, reflect correlates between contemporary climate and the social, cultural and colonial histories of cities in the Global South[45]. Inasmuch as urban planning has much inertia, especially in big cities, those choices might be correlated with climate because of the climatic correlates of colonial histories. On the other hand, it is also possible that these dynamics relate, in part, to the ways in which climate influences plant composition. This seems less likely given that under non-urban conditions vegetation cover (and hence cooling capacity) is normally positively correlated with MAT across the globe, opposite to our observed negative relationships for urban systems (Supplementary Fig. 2). However, it is possible that increased temperatures in cities due to the urban heat island effect may lead to temperature-vegetation cover-cooling capacity relationships that differ from those in natural environments[46-48]. Indeed, a recent study found that climate warming will put urban forests at risk, and the risk is disproportionately higher in the Global South[46]. Follow-up studies incorporating detailed sociocultural and ecological data are therefore needed to better unravel the underlying mechanisms.

**Potential of enhancing cooling and reducing inequality**

Can we reduce the inequality in cooling capacity and benefits that we have discovered among the world's largest cities? The good news is that our analyses suggest a huge potential to achieve this goal by optimizing urban green infrastructures. An obvious way is increasing urban green spaces. Assuming that maximum NDVI of a grid cell of a city sets the 'regional upper bounds' for NDVI of that same city as a whole, if we could systematically increase NDVI of all grid cells in each city to a level corresponding to the *median* of regional upper bounds (a conservative target, considering that the highest-level green space, e.g. Central Park of New York City, could hardly be expanded), while keeping cooling efficiency unchanged (see Methods), there would be a substantial amplification of cooling capacity by ~3.5 °C world-wide; and if a higher level of 90th percentile was reached, the gain of cooling capacity would be elevated to ~4.5 °C (Fig. 5a). The potential for cooling benefit is similar to that of cooling capacity (Extended Data Fig. 10). There is also potential to reduce urban temperatures if we can enhance cooling efficiency. However, the benefits of increases in cooling efficiency are modest (~1.5 °C increases at 90th percentile of regional upper bounds) when holding green space area constant. Ideally, if we could maximize both green space area and cooling efficiency (to 90 percentiles of their regional upper bounds respectively), we would yield increases in cooling capacity and benefit up to ~11 °C (9.7 °C in Global North and 11.5 °C in Global South), much higher than enhancing green space area or cooling efficiency alone (Fig. 5a, Extended Data Fig. 10a). Notably, such co-maximization of green space area and cooling efficiency would substantially reduce global inequality to Gini <0.1 (Fig. 5b, Extended Data Fig. 10b). Our analyses thus suggest that enhancing both green space 'quantity' and 'quality' can yield a synergistic effect leading to much larger gains than any single aspect alone.

**Outlook**

In summary, our results demonstrate clear inequality in the extent to which urban green spaces cool cities and their denizens between the Global North and South. Much attention has been paid to the global inequality of indoor heat adaptation arising from the inequality of resources (e.g., less affordable air conditioning and more frequent power shortage in the Global South)[49-52]. Our results suggest that the inequality of outdoor adaptation is at least equally concerning, particularly given that Global South urban populations are growing most rapidly is compounded by the fact that future temperature extremes are likely to be greatest in the Global South[16, 53].

Our results convey some good news in that there is a huge potential to strengthen the cooling capability of cities and to reduce inequalities in cooling capacities at the same time. Realizing this nature-based solution, however, requires smart planning strategies and advanced urban design and greening technologies[46]. Spatial planning of urban green space needs to consider not only the cooling amelioration effect, but also their multifunctional aspects that involve multiple ecosystem services, mental health benefits, accessibility, security, etc.[54-56]. In theory, a city can maximize its cooling while also maximizing density through the combination of high-density living, ground-level green spaces, and vertical and rooftop gardens (or even forests). In practice, the current cities with the most green space tend to be lower-density cities [57] (Supplementary Fig. 3). Innovation and implementation of new technologies that allow green space and high-density living to be combined have the potential to disconnect the negative relationship between green space area and population density[58, 59]. However, this appears to be a potential that has yet to be fully realized. Another dimension of green space area that deserves more attention is the geography of green space relative to denizens of cities. How best should we distribute green space in cities so as to maximize cooling efficiency[60, 61] and minimize within-city cooling inequality towards social equity[62]? It is also important to design and manage urban green space to itself be robust to future climate stress.

**Methods**

*Urban data*

We used the world population data from the World's Cities in 2018 Data Booklet[63] to select 509 major cities with population over 1 million (see Supplementary Table 1 for the complete list of the studied cities). For each selected city, we used the 2018 Global Artificial Impervious Area (GAIA) data at 30 m resolution[64] to determine its geographic extent. The derived urban boundary polygons thus encompass a majority of the built-up areas and urban residents. Our analyses on cooling amelioration were conducted based on MODIS remotely sensed data at 1 km resolution. We discarded the five cities with sizes <30 km² as they represent less than 30 MODIS grid cells of 1×1 km², and so were too small for us to estimate their cooling efficiency based on linear regression (see section below). We combined closely located cities that form contiguous urban areas or urban agglomerations, if their urban boundary polygons from GAIA merged (e.g., Phoenix and Mesa in the United States were combined). Our approach yielded 475 polygons, each representing a major urbanized area that were the basis for all subsequent analyses. Because large water bodies can exert substantial and confounding cooling effects, we excluded the 1×1 km² grid cells with water cover >20% using the European Space Agency (ESA) WorldCover data for 2020 at 10 m resolution[65].

*Quantifying the cooling effect*

As a first step, we calculated 'cooling efficiency' for each studied city within the GAIA-derived urban boundary. Cooling efficiency quantifies the extent to which a given area of green space in a city can reduce temperatures. It is a measure of the 'effectiveness' of urban green space in terms of heat amelioration. Cooling efficiency is typically measured by calculating the slope of the relationship between remotely-sensed land surface temperature (LST) and vegetation cover through ordinary least square regression; cooling efficiency is a proxy for the extent to which increases in vegetation per unit cover lead to decreases in temperature[34, 66]. It is known that cooling efficiency varies between cities. Influencing factors might include background climate[31], species composition[67], landscape configuration[60], and management practices[68]. However, the mechanism underlying the global pattern of cooling efficiency remains unclear.

We used MODIS satellite data to calculate the cooling efficiency of each studied city in the hottest month for that city. For LST we used the MODIS Aqua product (MYD11A1, 1 km resolution), which acquires remotely sensed images around 13:30 local time, representing daily maximum temperatures approximately. For each city we calculated mean LST in each month of 2018 to identify the hottest month, and then derived the hottest month LST. Correspondingly, we used the MODIS MYD13A2 product to calculate the mean NDVI for the hottest month. For each city, we conducted linear regression between LST and NDVI using the ordinary least square method. We calculated cooling efficiency as the slope. Before the analysis we discarded low-quality MODIS pixels of LST or NDVI, and then filtered out the pixels with NDVI <0 (normally less than 1% in a single city).

As a second step, we calculated the cooling capacity of each city. Cooling capacity is a positive function of the magnitude of cooling efficiency and the proportional area of green space in a city and is calculated based on MODIS NDVI and the derived cooling efficiency (Eq. 1, Supplementary Fig. 4):

$$CC = \frac{\sum_{i=1}^{n}(NDVI_i - NDVI_{min}) \times CE}{n} \quad (1)$$

where *CC* and *CE* are the cooling capacity and cooling efficiency of the focal city, respectively; $NDVI_i$ is NDVI for 1-km grid cell *i*; $NDVI_{min}$ is the minimum NDVI across the city; and *n* is the total number of 1 km grid cells within the city. Local cooling capacity for each grid cell *i* (Fig. 1, Extended Data Fig. 8) can be derived in this way as well (Supplementary Fig. 3). For a particular city, cooling capacity may be dependent on the spatial configuration of its land use/cover[60, 61], but here we condensed cooling capacity to city average, thus did not take into account these local-scale factors.

As a third step, we calculated the cooling benefit realized by an average urban resident (cooling benefit in short) in each city. Cooling benefit depends not only on the cooling capacity of a city, but also on where people live within a city relative to greener or greyer areas of the city. For example, cooling benefits in a city might be low even if the cooling capacity is high if the green parts and the dense-population parts of a city are inversely correlated. Here, we are calculating these averages while aware that in any particular city that the exposure of a particular person will depend on the distribution of green space in a city, and the occupation, movement trajectories of a person, etc. On the scale of a city, we calculated cooling benefit following a previous study[69], that is, simply adding a weight term of population size per 1-km grid cell into cooling capacity in Eq. (1):

$$CB = \frac{\sum_{i=1}^{n}(NDVI_i - NDVI_{min}) \times CE \times pop_i}{\sum_{i=1}^{n} pop_i} \quad (2)$$

Where *CB* is cooling benefit, $pop_i$ is the number of people within grid cell *i*. The population data were obtained from the 1-km resolution Gridded Population of World Version 4, Revision 11 (GPWv411) of 2015[70]. Local cooling benefit for a given grid cell *i* can be calculated in a similar way, i.e., local cooling capacity multiplied by a weight term of local population density relative to mean population density. Local cooling capacity and benefit were mapped for example cities for the purpose of illustrating the effect of population spatial distribution (Fig. 1 and Extended Data Fig. 8), but their patterns were not examined here.

We conducted univariate analyses to examine if and to what extent cooling efficiency and cooling benefit are shaped by essential natural and socioeconomic factors, including background climate (mean annual temperature[71] and precipitation[72]), topography (elevation range[73]), and GDP per capita[74], with city size (geographic extent) corrected for. We did not include humidity because it is strongly correlated with temperature and precipitation, causing serious multi-collinearity problem. We used piecewise structural equation modeling to test the direct effects of these factors and indirect effects via influencing cooling efficiency and vegetation cover (Fig. 4, Extended Data Fig. 9). To account for the potential influence of spatial autocorrelation, we used spatially autoregressive models (SAR) to test for the robustness of the observed effects of natural and socioeconomic factors on cooling capacity and benefit (Supplementary Fig. 5).

*Testing for robustness*

We conducted five additional analyses to test for robustness:

(1) We unfold the approach above using MODIS Enhanced Vegetation Index (EVI) to check if the results are robust to different spectral greenness indicators (Supplementary Fig. 6).

(2) We looked at mean hottest-month LST and NDVI within 5 years (2014-2018) to check the consistency between the results based on relatively short (1 year) vs. long (5-year average) time periods (Supplementary Fig. 7).

(3) We quantified the cooling amelioration based on *relative* temperature cooled down by urban green space (Supplementary Figs. 8, 9). This quantification is based on the consideration that the amelioration effect perceived by humans is also dependent on background temperatures. For example, while Khartoum (Sudan) and London (UK) present a similar cooling capacity around 2 °C, their difference in hottest monthly temperature is huge: 34 °C in Khartoum vs 18 °C in London. Therefore, the same level of 'absolute' cooling capacity in Khartoum is more critical to human health than in London. The relative cooling capacity $CB_{Relative}$ is calculated as:

$$CB_{Relative} = \frac{CB}{T_{max} - T} \qquad (3)$$

Where $T_{max}$ is the maximum of hottest monthly temperature among all human settlements (population density >10) in 2018, i.e., 45 °C; $T$ is the mean hottest monthly temperature for the focal city.

(4) For the calculation of cooling benefit, we considered different spatial scales of human accessibility to green space: assuming population in each 1×1 km² grid cell could access to green space within neighborhoods of certain extents, we calculated cooling benefit by replacing $NDVI_i$ in Eq. (2) with mean NDVI within the 3×3 km² and 5×5 km² extents centered at the focal grid cell (Supplementary Figs. 10, 11).

(5) Considering cities may vary in minimum NDVI, we assessed if this variation could affect resulting cooling capacity patterns. To this end, we calculated cooling capacity for each studied city using NDVI = 0 as the reference (i.e., using NDVI = 0 instead of minimum NDVI in Supplementary Fig. 4b), and correlated it with that using minimum NDVI as the reference (Supplementary Fig. 12).

We obtained consistent results from these five robustness analyses.

*Quantifying between-city inequality*

Inequalities in access to the benefits of green space in cities exist *within* cities, as is increasingly well-documented[34]. Here, we focus instead on the inequalities *among* cities. We used the Gini coefficient to measure the inequality in cooling capacity and cooling benefit between all studied cities across the globe as well as between Global North or South cities. We calculated Gini using the population-density weighted method (Fig. 5b), as well as the unweighted and population-size weighted methods (Supplementary Figs. 1, 13).

*Estimating the potential towards more effective and equal cooling amelioration*

We estimated the potentials of enhancing cooling amelioration based on the assumptions that urban green space quality (cooling efficiency) and quantity (NDVI) can be increased to different levels, and that relative spatial distributions of green space and population can be idealized (so that their spatial matches can maximize cooling benefit). We assumed that climate conditions act as the constraint of vegetation cover and cooling efficiency. We calculated the maximum NDVI for each city within a given climate region, assembling these values as 'upper bounds'. For each city, we generated a potential NDVI distribution where all grid cells reach the $50^{th}$, $60^{th}$, $70^{th}$, $80^{th}$, or $90^{th}$ percentile of the upper bound values. NDVI values below these percentiles were increased, whereas those above these percentiles remained unchanged. We used the Köppen climate classification system[75], and tropical, arid, temperate, and continental climate regions were involved for all studied cities. We calculated potential cooling capacity and cooling benefit based on these potential NDVI maps ('Fixed green space area' in Fig. 5). We then calculated the potentials if cooling efficiency of each city can be enhanced to 50-90th percentile across all cities within the corresponding biogeographic region ('Fixed cooling efficiency' in Fig. 5). We also calculated the potentials if both NDVI and cooling efficiency were enhanced ('Enhancing both' in Fig. 5) to a certain corresponding level (i.e., $i$th percentile NDVI + $i$th percentile cooling efficiency). We examined if there are additional effects of idealizing relative spatial distributions of urban green spaces and humans on cooling benefit. To this end, the pixel values of NDVI or population amount remained unchanged, but their one-to-one correspondences were based on their ranking: the largest population correspond to the highest NDVI, and so forth. Under each scenario, we calculated cooling capacity and cooling benefit for each city, and the between-city inequality was measured by the Gini coefficient.

The spatial data analyses were conducted using the cloud computing platform Google Earth Engine and ArcGIS v10.2 software. The statistical analyses were conducted using R v4.3.0[76], and *piecewiseSEM* packages v2.1.2[77].

**Acknowledgements**

We thank all the data providers. We thank Marten Scheffer for valuable discussion. This work is supported by the National Natural Science Foundation of China (Grant No. 32061143014) and the Open Fund for Key Lab. of Land Degradation and Ecological Restoration in northwestern China of Ningxia University (Grant No. LDER2023Z01). JCS was supported by Center for Ecological Dynamics in a Novel Biosphere (ECONOVO), funded by Danish National Research Foundation (grant DNRF173) and his VILLUM Investigator project "Biodiversity Dynamics in a Changing World", funded by VILLUM FONDEN (grant 16549). T.M.L. and J.F.A. are supported by the Open Society Foundations (OR2021-82956). T.M.L. is supported by a Turing Fellowship.

**Author contributions**

Y.L., S.N.T., R.R.D. and C.X. designed the study. Y.L. and S.N.T. performed the analyses with input from J.-C.S., W.Z., K.Z., J.F.A., T.M.L, R.D.D. and C.X. Y.L. and S.N.T. produced the figures with input from R.D.D., J.-C.S. and C.X. Y.L., S.N.T., R.R.D. and C.X. wrote the paper with input from J.-C.S., W.Z., K.Z., J.F.A., and T.M.L.

**Competing interests**

The authors declare no competing interests.

**Data and code availability**

City population statistics data is collected from the Population Division of the Department of Economic and Social Affairs of the United Nations (https://www.un.org/development/desa/pd/content/worlds-cities-2018-data-booklet). Global urban boundaries from GAIA data is available from FROM-GLC group of Tsinghua University (http://data.ess.tsinghua.edu.cn/gub.html). Global water data is derived from European Space Agency (ESA) WorldCover 10 m 2020 product (https://zenodo.org/record/5571936). Land surface temperature (LST) data from MODIS Aqua product (MYD11A1) is available at https://doi.org/10.5067/MODIS/MYD11A1.061. NDVI dataset from MYD13A2 is available at https://doi.org/10.5067/MODIS/MYD13A2.006. Population data is derived from the GPWv4, which is publicly available from Center for International Earth Science Information Network (CIESIN) (https://doi.org/10.7927/H4PN93PB). Mean annual temperature data is calculated from ERA5-Land Monthly Aggregated dataset (https://doi.org/10.24381/cds.68d2bb30). Mean annual precipitation data is calculated from TerraClimate (Monthly Climate and Climatic Water Balance for Global Terrestrial Surfaces, University of Idaho) (https://doi.org/10.1038/sdata.2017.191). Topography data from MERIT DEM (Multi-Error-Removed Improved-Terrain DEM) product is available at https://doi.org/10.1002/2017GL072874. GDP from Gross Domestic Product and Human Development Index dataset is available at https://doi.org/10.5061/dryad.dk1j0. City building volume data from Global 3D Building Structure is available at https://doi.org/10.34894/4QAGYL. The codes used for the analyses will be publically available at Figshare.

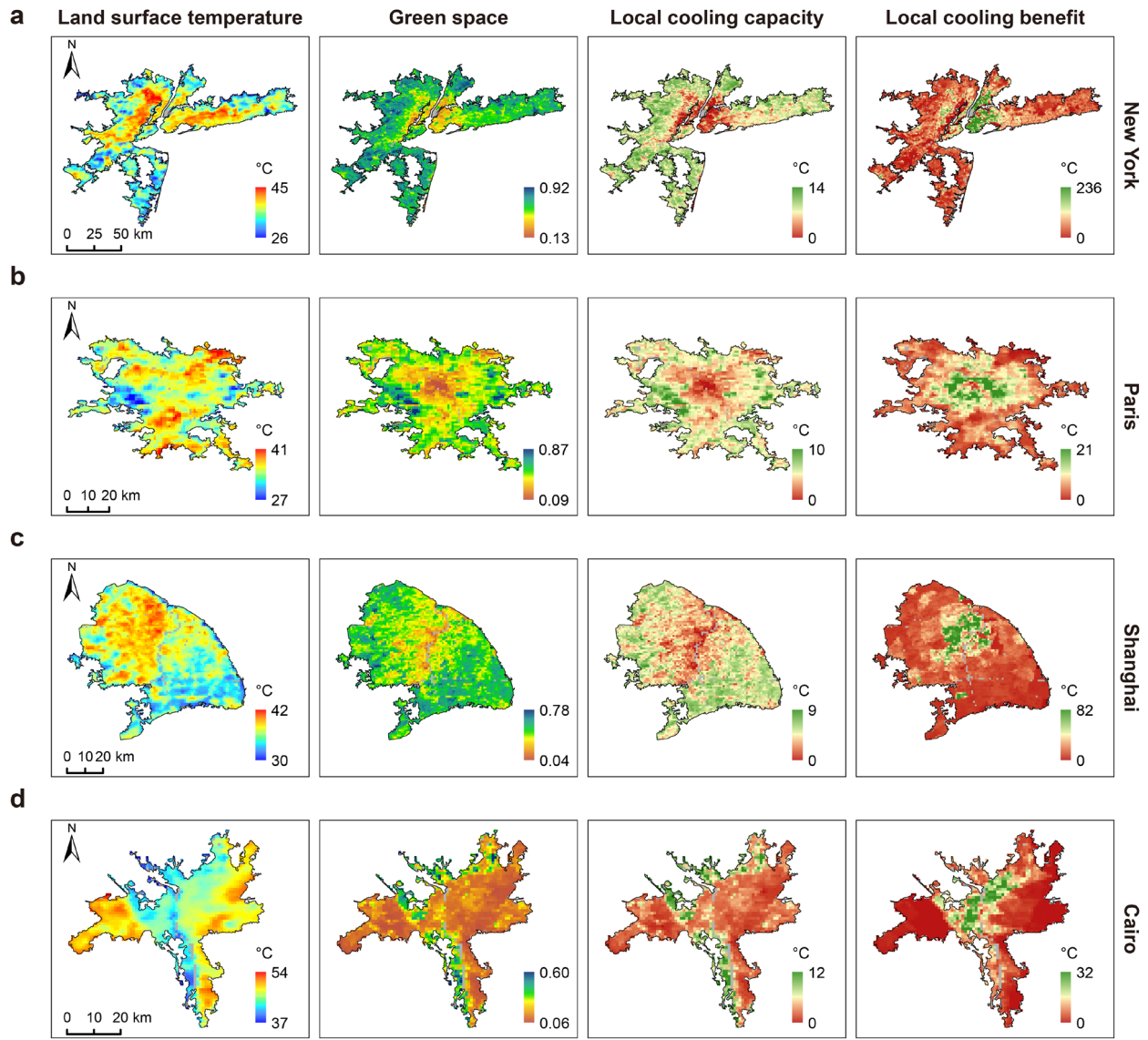

**Fig. 1. Distributions of city-scale land surface temperature, green space (MODIS NDVI), and derived local cooling capacity and local cooling benefit exemplified by four mega cities across different continents. *a*,** New York-Newark, US. ***b*,** Paris, France. ***c*,** Shanghai, China. ***d*,** Cairo, Egypt. In densely populated parts of cities, local cooling capacity tends to be lower due to reduced green space area, whereas local cooling benefit (local cooling capacity multiplied a weight term of local population density relative to city mean) tends to be higher as more urban residents can receive cooling amelioration. Note, the larger geographic scale for New York City and its relatively homogenous green space (outside of the very highest density parts of the city, such as Manhattan with very high local cooling benefit), the strong geographic pattern of green

space and its benefits in Shanghai, and the near absence of green space in Cairo.

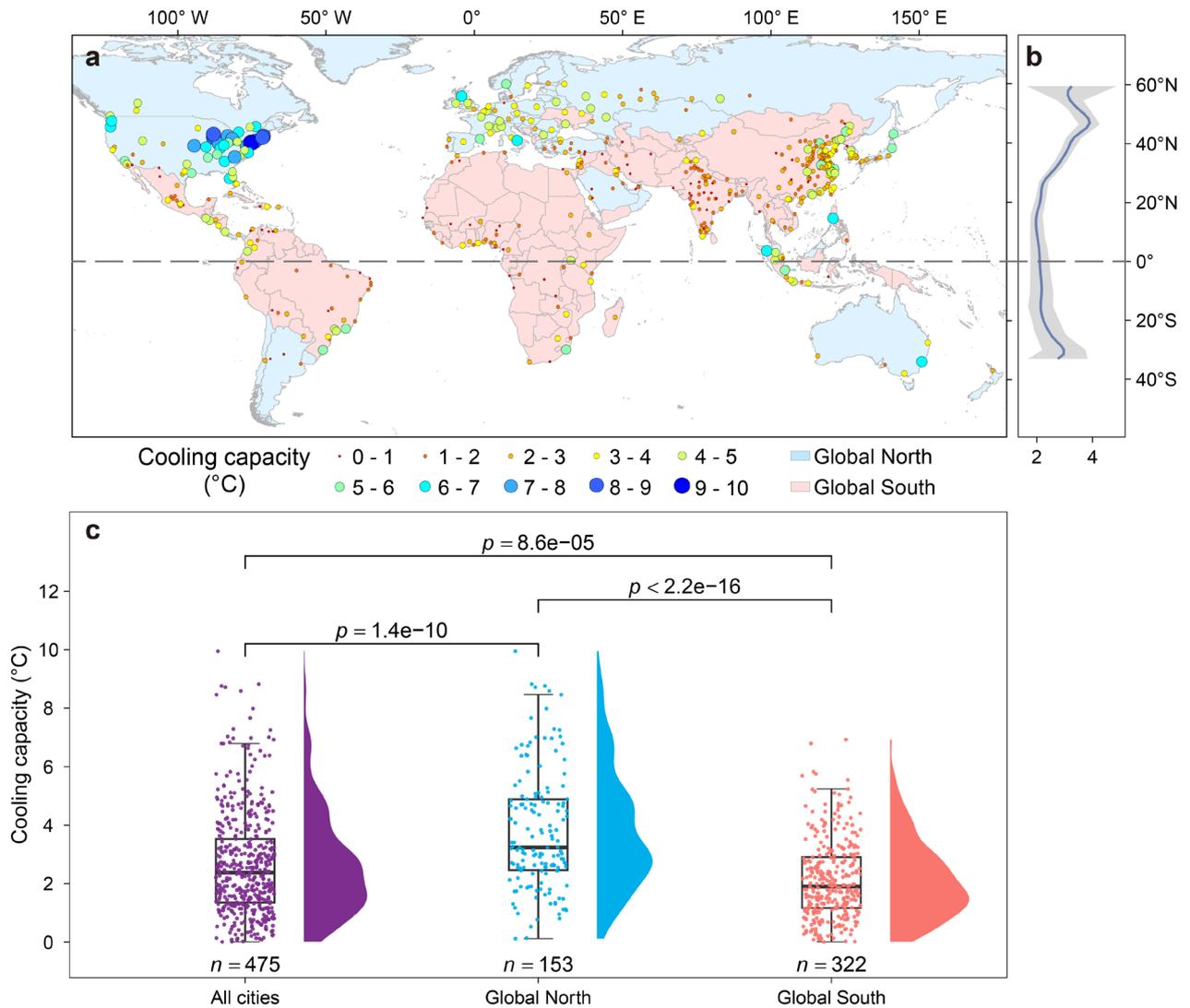

**Fig. 2. Global pattern of cooling capacity.** *a,* Global distribution of cooling capacity for the 475 major urbanized areas. *b,* Latitudinal pattern of cooling capacity. *c,* Cooling capacity difference between the Global North and South cities. The cooling capacity offered by urban green infrastructure evinces a latitudinal pattern wherein lower-latitude cities have weaker cooling effects (*b,* cubic-spline fitting with 95% confidence interval is shown), representing a significant inequality between Global North and South countries: city-level cooling capacity for Global North cities are about two-fold higher than in Global South cities (*c*). The tails of the cooling capacity distributions are truncated at zero as all cities have positive values of cooling capacity. Notice that no cities in the Global South have a cooling capacity greater than 7 °C (*c*). This is because no cities in the Global South have proportional green space areas as great as those seen in the Global

North (see also Fig. 4b). A similar pattern is found for cooling benefit (Extended Data Fig. 3). The non-parametric Wilcoxon test was used for statistical comparisons.

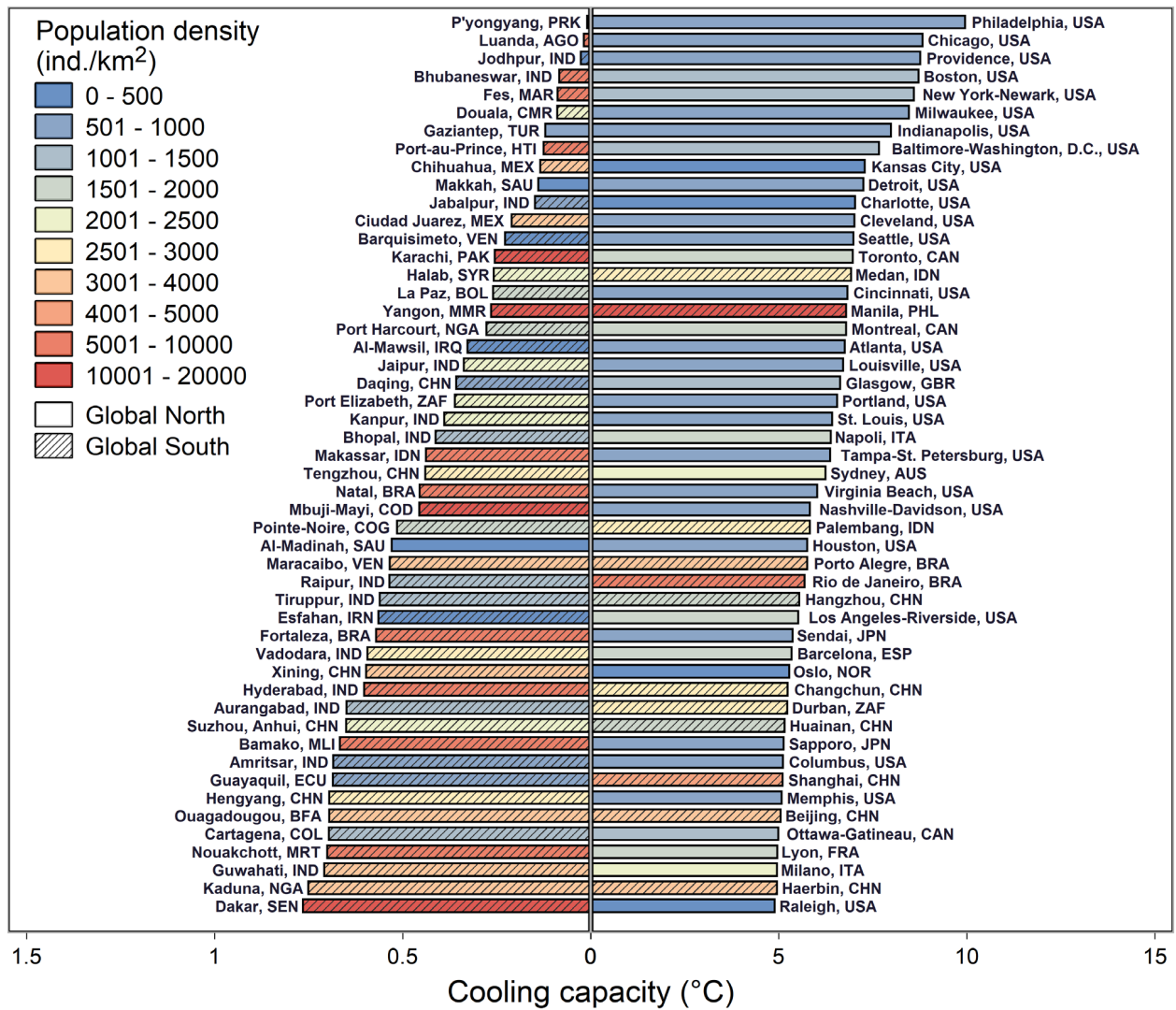

**Fig. 3. Contrast between the 50 cities with highest (right hand bars) and those with lowest (left hand bars) cooling capacities.** The axes on the right are an order of magnitude greater than those on the left, such that the cooling capacity of Philadelphia in the United States is about 1000-fold greater than that of P'yongyang (North Korea) and 500-fold greater than that of Luanda (Angola). The cities presenting lowest cooling capacities are most associated with Global South cities at higher population densities. Interestingly, the cities with the highest cooling capacity tend to be relatively large (and those with the lowest cooling capacity relatively small, Extended Data Fig. 5). This is hopeful in that it suggests that high population densities, even those associated with cities such as Beijing, Shanghai and New York, are reconcilable with large quantities of green space.

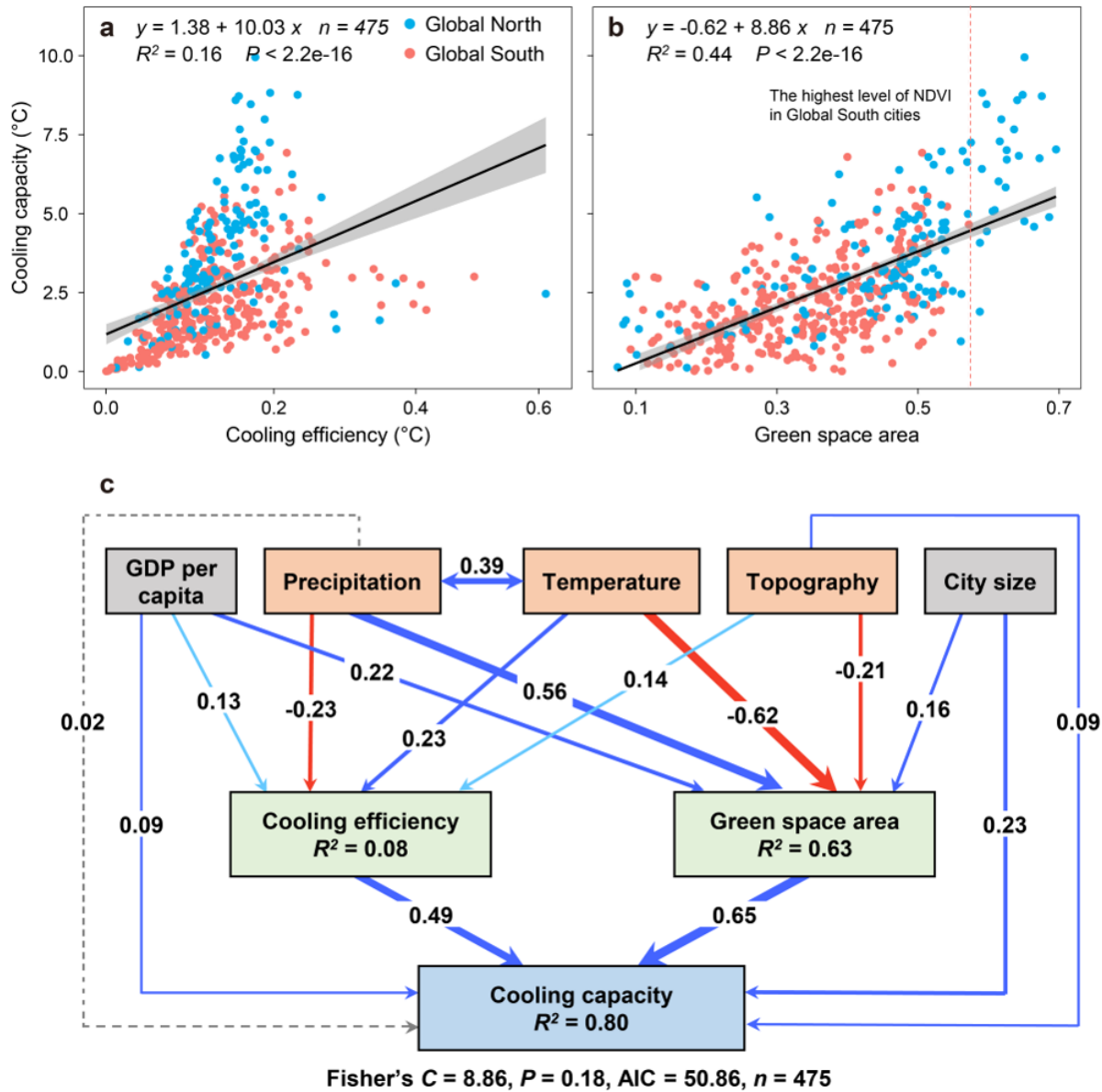

**Fig. 4. Cooling capacity is significantly correlated with cooling efficiency and urban green space area, which are jointly shaped by natural and social economic factors.** *a,* Relationship between cooling efficiency and cooling capacity. *b,* Relationship between urban green space area (measured by mean MODIS NDVI in the hottest month of 2018) and cooling capacity. Note that the highest level of urban green space area in the Global South cities is much lower than that in the Global North (dashed line in *b*). *c,* A piecewise structural equation model based on assumed direct and indirect (through influencing cooling efficiency and urban green space area) effects of essential natural and socioeconomic factors on cooling capacity. Mean annual temperature and

precipitation, and topography (elevation range) are selected to represent basic background natural conditions; GDP per capita is selected to represent basic socioeconomic conditions. Spatial extent of built-up areas is included to correct for city size. A bi-directional relationship (correlation) is fitted between mean annual temperature and precipitation. Red and blue solid arrows indicate significantly negative and positive coefficients with $p \leq 0.05$, respectively (dark blue: $p<0.001$; light blue: $0.001 \leq p \leq 0.05$). Gray dashed arrows indicate $p >0.05$. Arrow width illustrate effect size. Similar relationships are found for cooling benefit realized by an average urban resident (see Extended Data Fig. 9). We obtained $p$ values based on a two-sided test for the estimated slopes.

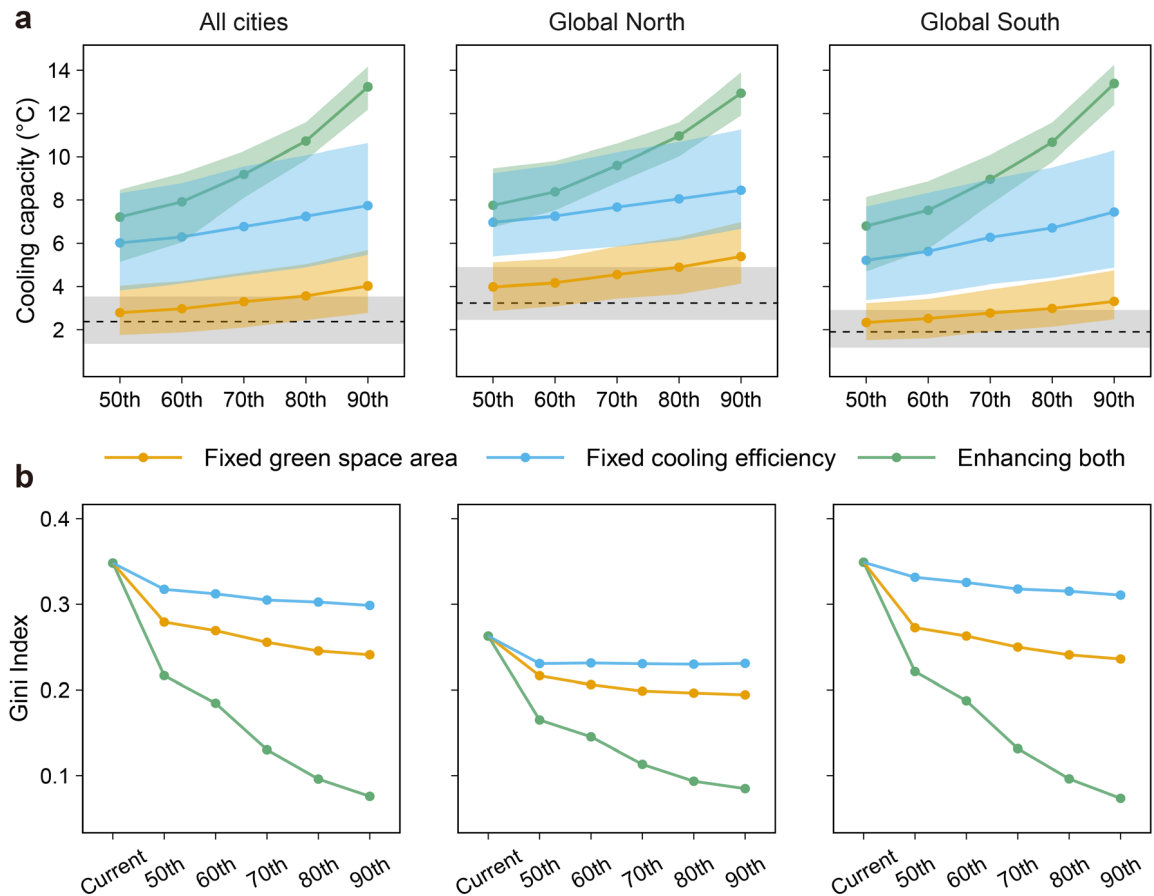

**Fig. 5. Estimated potential of enhancing cooling capacity and reducing its inequality.** *a,* The potential of enhancing cooling capacity via either enhancing cooling efficiency (holding green space area fixed, yellow curves), or increasing urban green space area (holding cooling efficiency fixed, blue curves) alone is much lower than that of enhancing both cooling efficiency and urban green space area (green curves). The dashed horizontal lines and the gray bands represent median and 25-75[th] percentiles of cooling capacity of current cities. The colored curves (median) and bands (95% confidence interval) represent the effects of enhancing cooling efficiency (yellow), urban green space area (blue), or both (green) to 50-90[th] percentile of their assumed regional upper bounds, respectively (see Methods). *b,* The potential of reducing cooling capacity inequality is also higher when enhancing both cooling efficiency and urban green space area. Gini index weighted by population density is used to measure inequality. Similar results were found for cooling benefit (Extended Data Fig. 10).

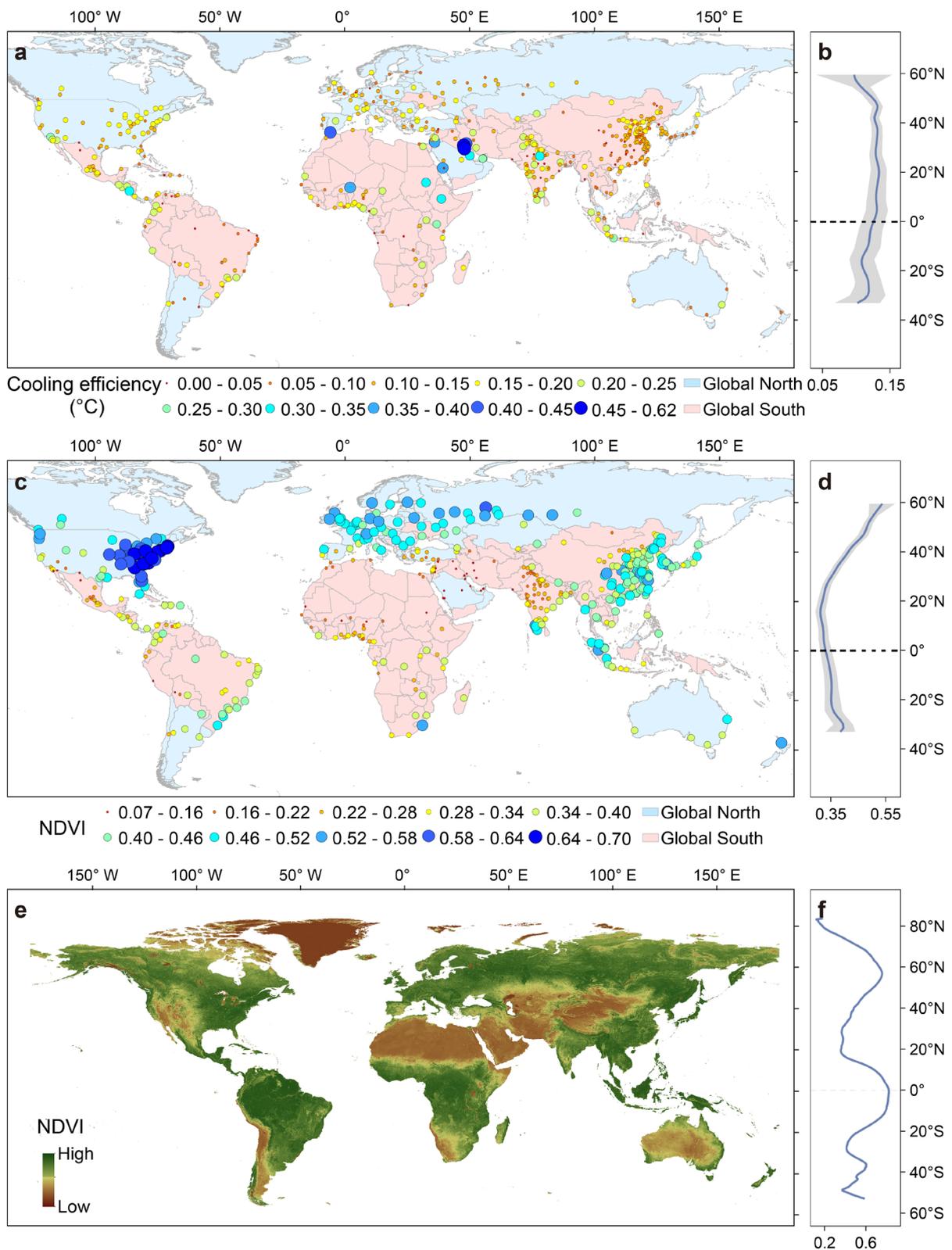

**Extended Data Fig. 1. Global patterns of cooling efficiency and urban green space area across 475 major urban areas.** *a-b,* Cooling efficiency of urban green space is lower around 20°S. *c-d,* Urban green space area (measured by mean within-city MODIS NDVI in the hottest

month of 2018) is higher in the northern hemisphere, especially in Northern America and Europe.

*e-f,* Global vegetation cover (measured by monthly maximum MODIS NDVI in 2018) is peaked at the equator, presenting a different pattern than in urban areas.

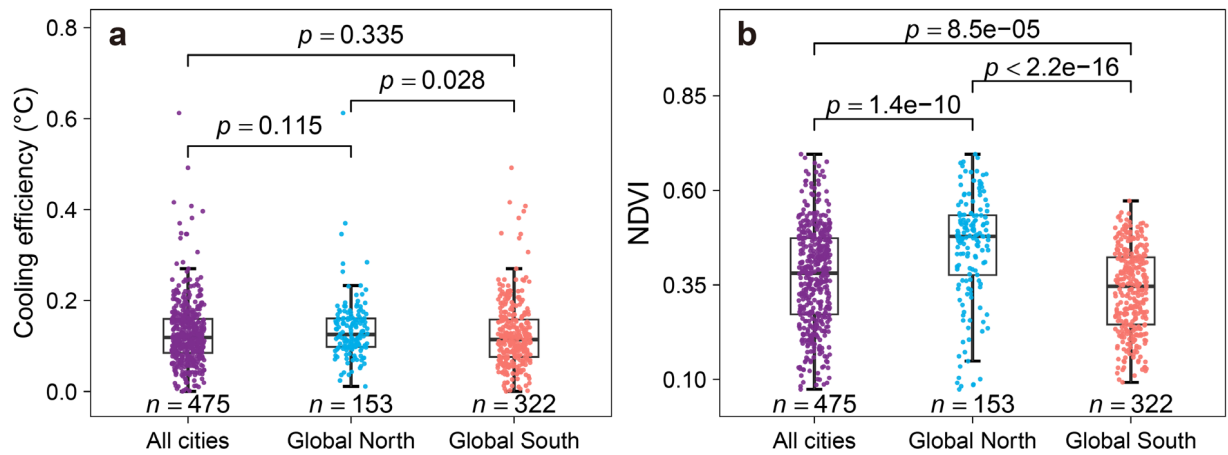

**Extended Data Fig. 2. Differences of cooling efficiency and urban green space area between the Global North and South.** *a,* Cooling efficiency. *b,* Urban green space area (measured by mean within-city MODIS NDVI in the hottest month of 2018). The mean cooling efficiency of all studied 475 major urbanized areas is 0.13 °C, i.e., a 1% increase in urban green space can reduce land surface temperature by 0.13 °C. The Global North cities have a slightly higher level of mean cooling efficiency of 0.14±0.07 °C than the Global South of 0.13±0.07 °C (*a*). The mean urban green space area (measured by NDVI) of the Global North cities is significantly higher than the Global South (0.45±0.14 vs 0.34±0.11). The mean NDVI for all studied cities is 0.37. Mean ± s.d. are indicated in brackets. The non-parametric Wilcoxon test was used for statistical comparisons

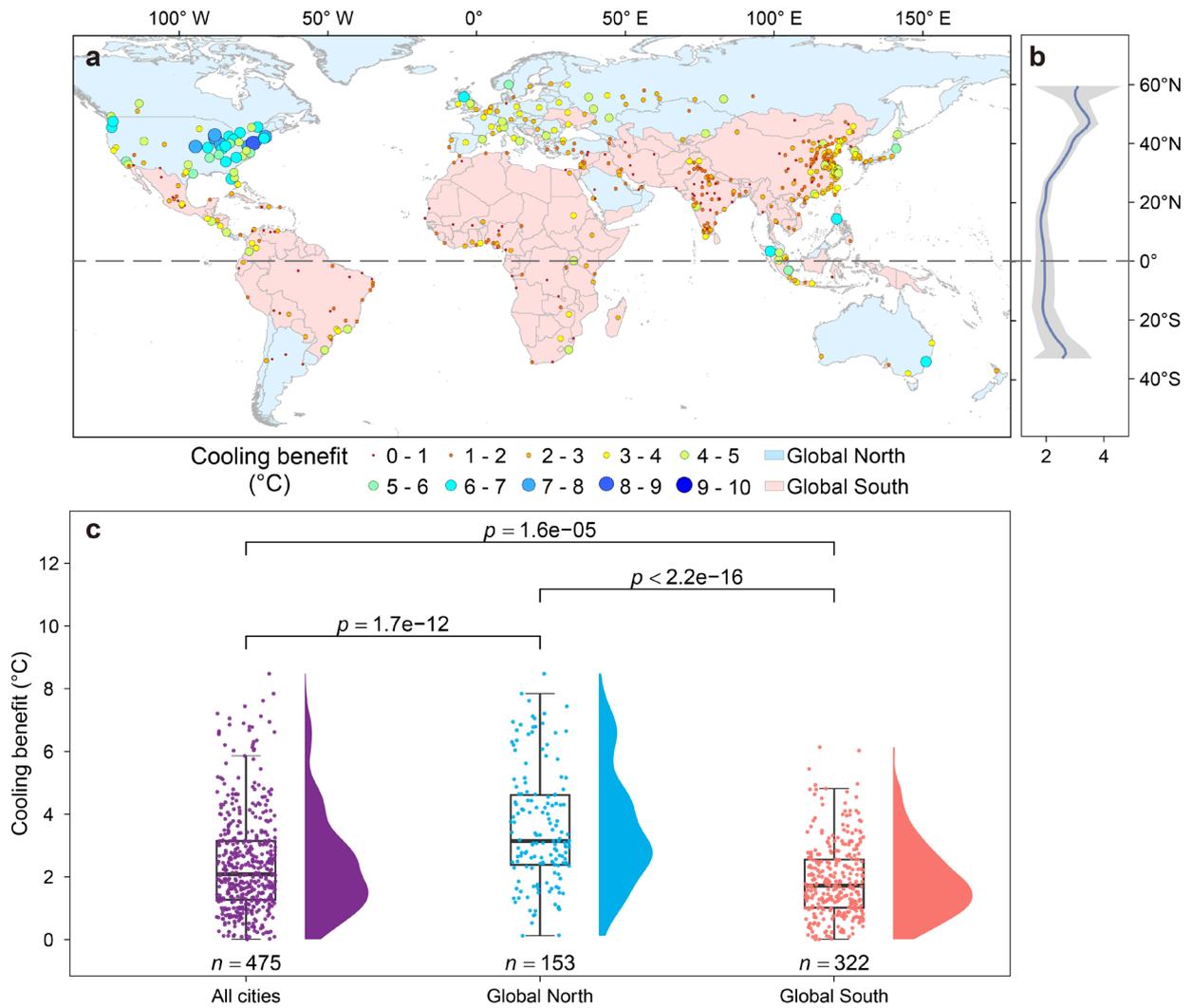

**Extended Data Fig. 3. Cooling benefit realized by an average urban resident presents a similar global pattern as cooling capacity.** *a*, Global distribution of cooling benefit for the 475 major urbanized areas. *b*, Latitudinal pattern of cooling benefit. *c*, Cooling benefit of the Global North cities is about two-fold higher than the Global South. The non-parametric Wilcoxon test was used for statistical comparisons.

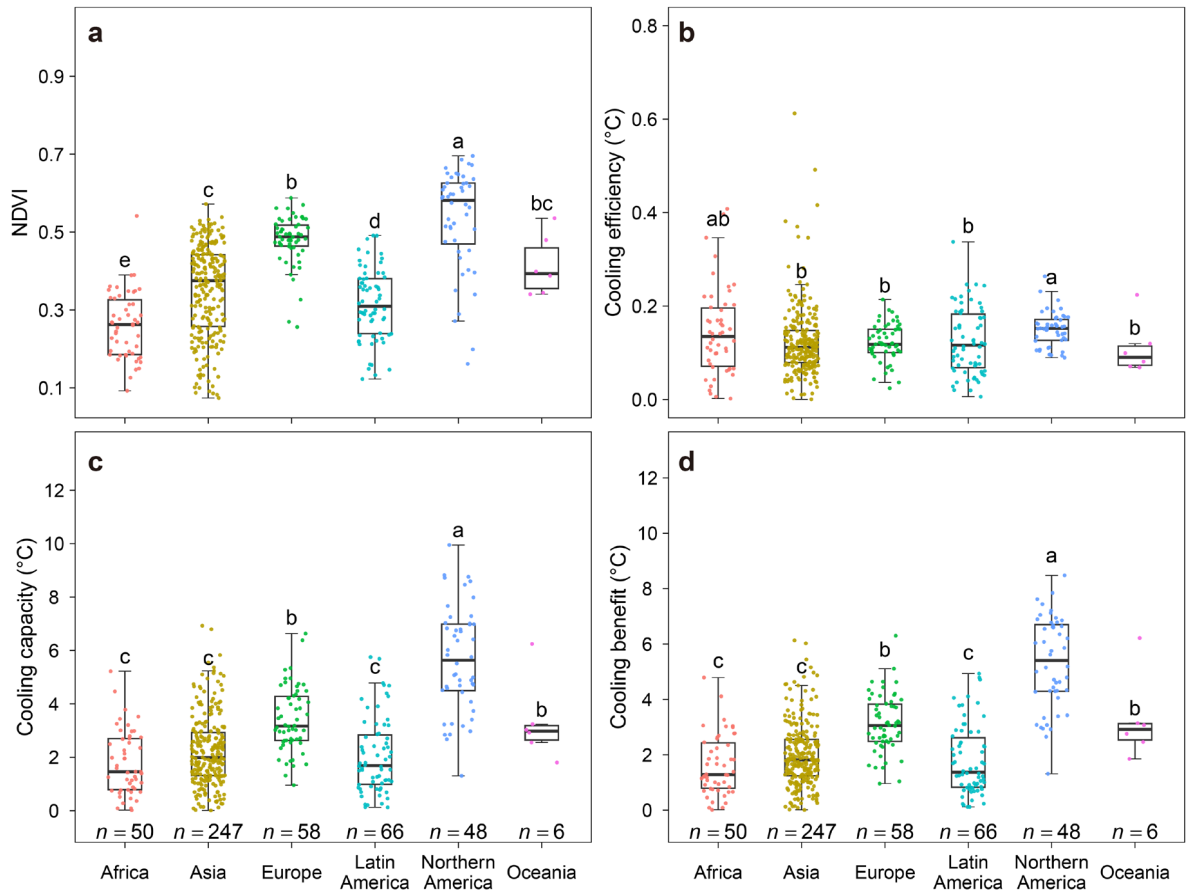

**Extended Data Fig. 4. Continental differences of urban green space and its cooling capabilities.** *a,* Urban green space area (measured by mean within-city MODIS NDVI in the hottest month of 2018). *b,* Cooling efficiency. *c,* cooling capacity. *d,* Cooling benefit realized by an average urban resident. Urban green space area varies significantly among the different continents: Northern America (including Canada and the US while excluding the Central American countries) present the highest NDVI (0.53±0.13), followed by Europe (0.48±0.06), Oceania (0.41±0.08), Asia (0.35±0.12), Latin America (0.31±0.09), and Africa (0.26±0.09) (*a*). Cooling capacity and cooling present a similar pattern of continental differences. Cooling efficiency presents a different order: Northern America (0.15±0.13 °C), Africa (0.14±0.09 °C), Latin America (0.13±0.09 °C), Asia (0.12±0.12 °C), Europe (0.12±0.06 °C), and Oceania (0.11±0.08 °C). Mean ± s.d. are indicated in brackets. The non-parametric Wilcoxon test was used for statistical comparisons. Different letters above boxes indicate significant differences with *p* <0.05.

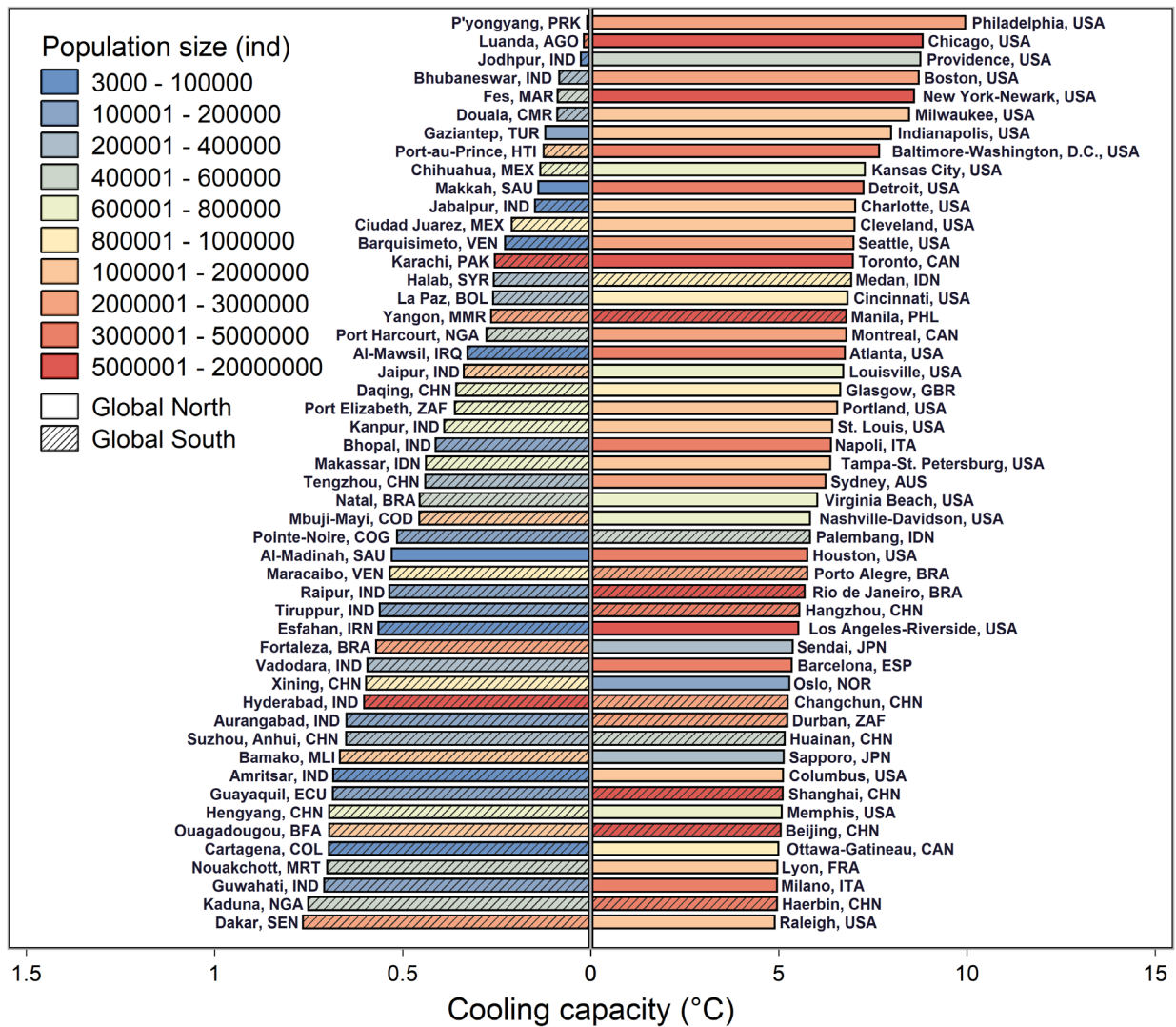

**Extended Data Fig. 5. Ranking of the 50 most (right hand bars) and least (left hand bars) effective cities in terms of cooling capacity with population size.** Cities with high cooling capacity typically accommodate larger populations. For example, population size in Philadelphia (the US) where cooling capacity is the highest is eight-fold greater than in P'yongyang (North Korea) where cooling capacity is the lowest. Note that here population size was calculated within the artificial impervious areas rather than administrative boundaries.

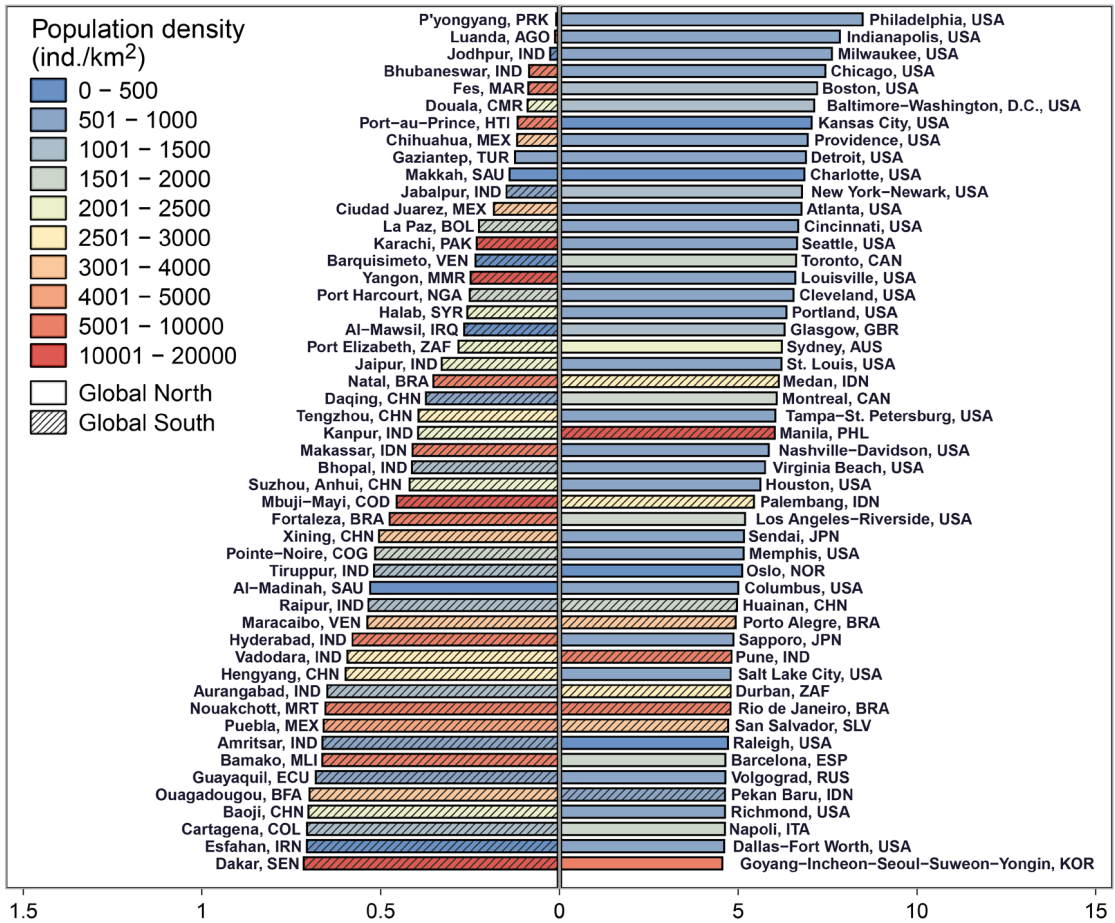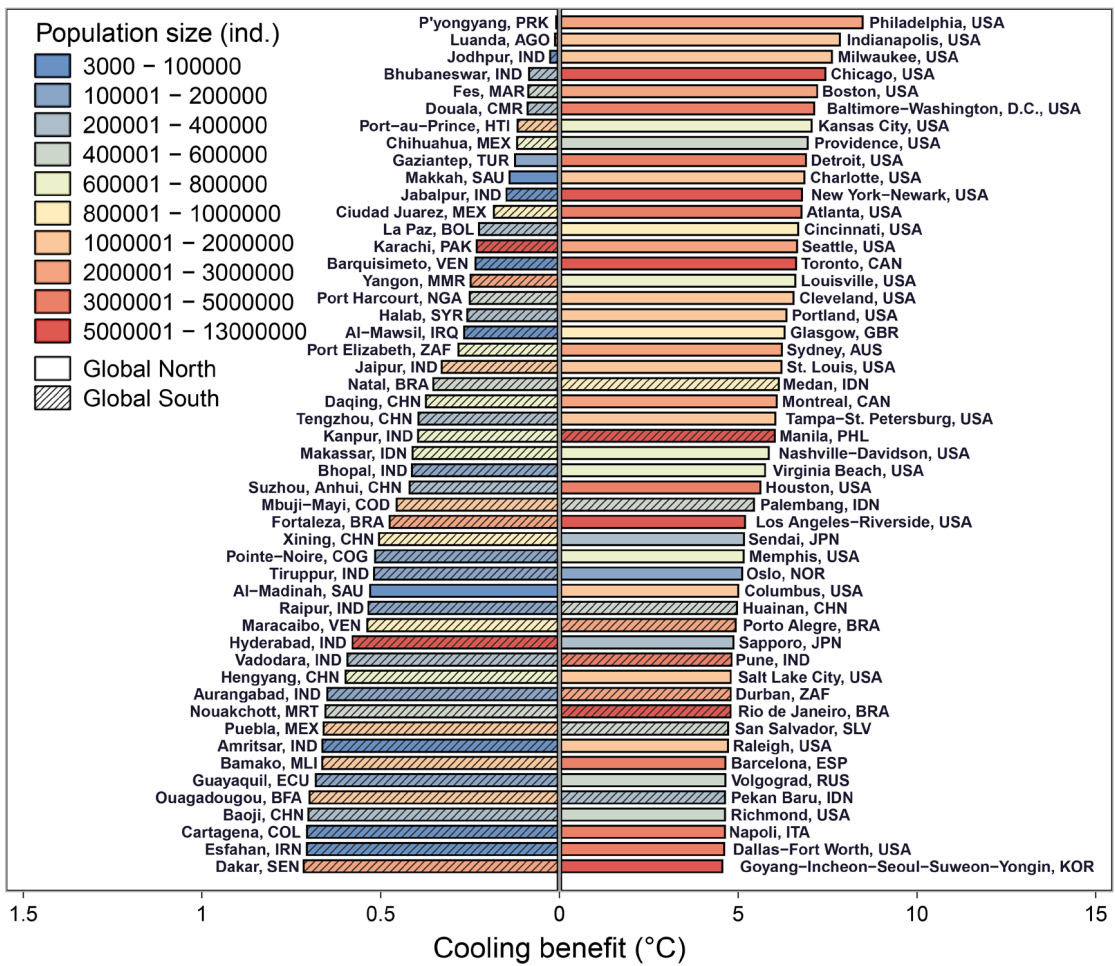

**Extended Data Fig. 6. Contrast between the 50 cities with highest (right hand bars) and those with lowest (left hand bars) cooling benefit realized by an average urban resident within a city. *a,*** Population density per city shown in color gradient. ***b,*** population size per city shown in color gradient. Note that the axes on the right are an order of magnitude greater than those on the left. The cities are predominantly located in the Global North with lower population densities and larger population sizes. Note that here population size was calculated within the polygons of artificial impervious areas rather than administrative boundaries.

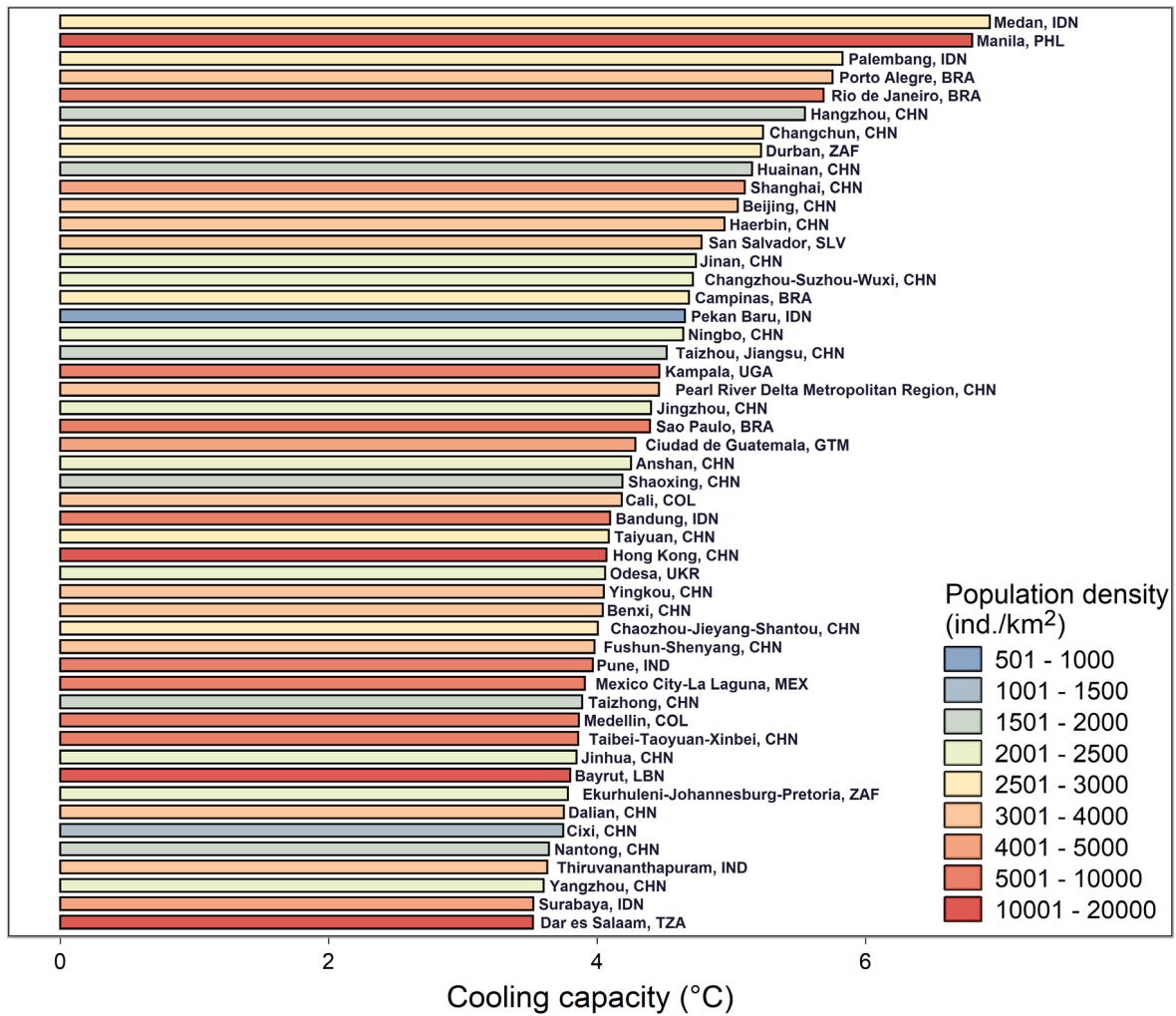

**Extended Data Fig. 7. Ranking of the 50 most effective cities in terms of cooling capacity in association with population density.** About 70% of these cities are located in China (27), Indonesia (5), and Brazil (4). Medan (Indonesia) has the highest cooling capacity (~7 °C) among all Global South cities. This is likely due to a combination of high overall green space area (mean NDVI of 0.51) and cooling efficiency (0.22 °C, potentially influenced by background climate and geographical location).

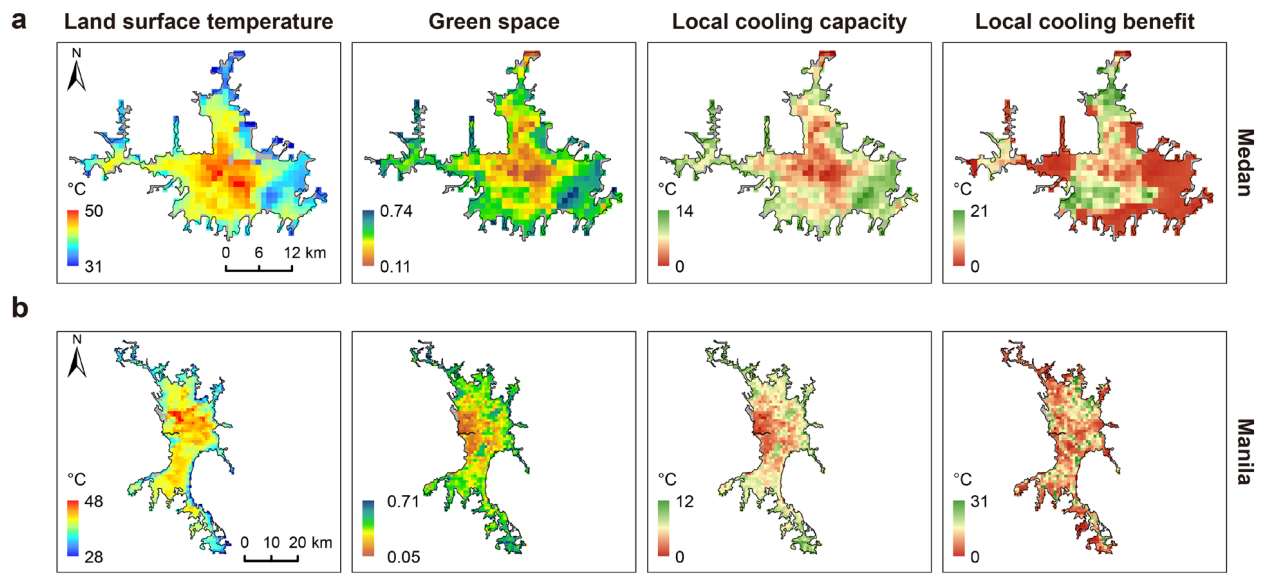

**Extended Data Fig. 8. Distributions of city-scale land surface temperature, green space (MODIS NDVI), and cooling capability of two Global South cities with high cooling capacity.** These cities exhibit higher cooling capacity due to both high urban green space area and cooling efficiency. *a,* Medan, Indonesia. Medan stands out among Global South cities with the highest cooling capacity of ~7 °C, which is 3.5-fold greater than the mean Global South cooling capacity. Medan presents mean MODIS NDVI of 0.51 in the hottest month of 2018 (ranking the 14th among the 322 Global South cities), and cooling efficiency of 0.22 °C (ranking the 29th among the 322 Global South cities). *b,* Manila, the Philippines. Manila presents a cooling capacity of 6.8 °C (ranking the 2nd among the Global South cities), mean MODIS NDVI of 0.40 (ranking the 103rd among the Global South cities), and cooling efficiency of 0.18 °C (ranking the 57th among the Global South cities). This example indicates that a combination of even moderate levels of urban green space area and cooling efficiency could yield high cooling capacity. For most Global South cities, their cooling capacity is limited by either relatively small green space area or relative low cooling efficiency.

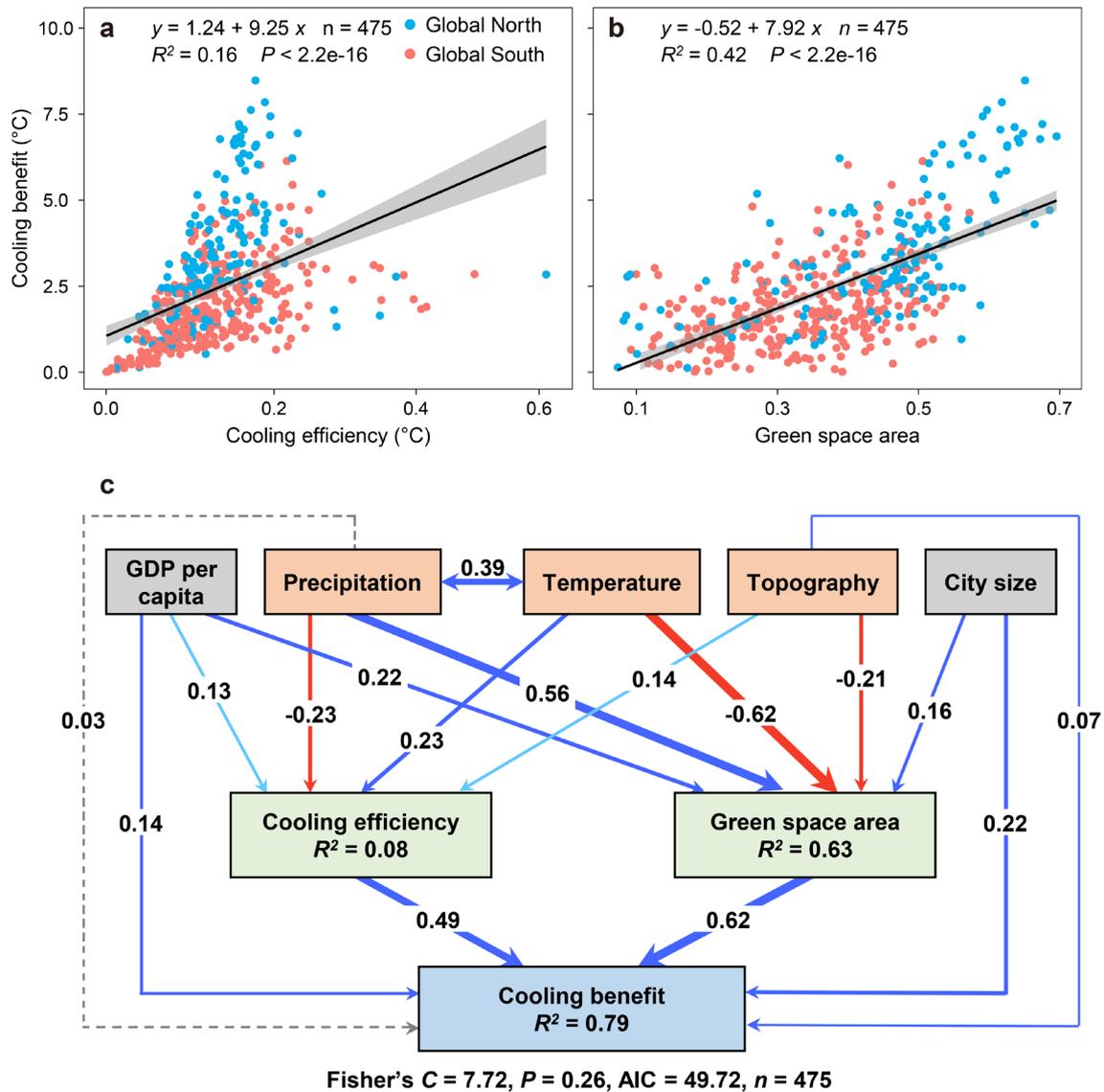

**Extended Data Fig. 9. Cooling benefit realized by an average urban resident is significantly correlated with cooling efficiency and urban green space area, which are jointly shaped by natural and social economic factors.** *a,* Relationship between cooling efficiency (the extent to which a given area of green space in a city can reduce temperatures, see Methods) and cooling benefit. *b,* Relationship between urban green space area (measured by mean MODIS NDVI in the hottest month of 2018) and cooling benefit. *c,* A piecewise structural equation model based on assumed direct and indirect (through influencing cooling efficiency and urban green space area) effects of essential natural and socioeconomic factors on cooling benefit. Mean annual temperature and precipitation, and topography (elevation range) are selected to represent basic

background natural conditions; GDP per capita is selected to represent basic socioeconomic conditions. Spatial extent of built-up areas is included to correct for city size. A bi-directional relationship (correlation) is fitted between mean annual temperature and precipitation. Red and blue solid arrows indicate significantly negative and positive coefficients with $p \leq 0.05$, respectively (dark blue: $p<0.001$; light blue: $0.001 \leq p \leq 0.05$). Gray dashed arrows indicate $p >0.05$. Arrow width illustrate effect size. Similar relationships are found for cooling capacity (Fig. 4). We obtained *P* values based on a two-sided test for the estimated slopes.

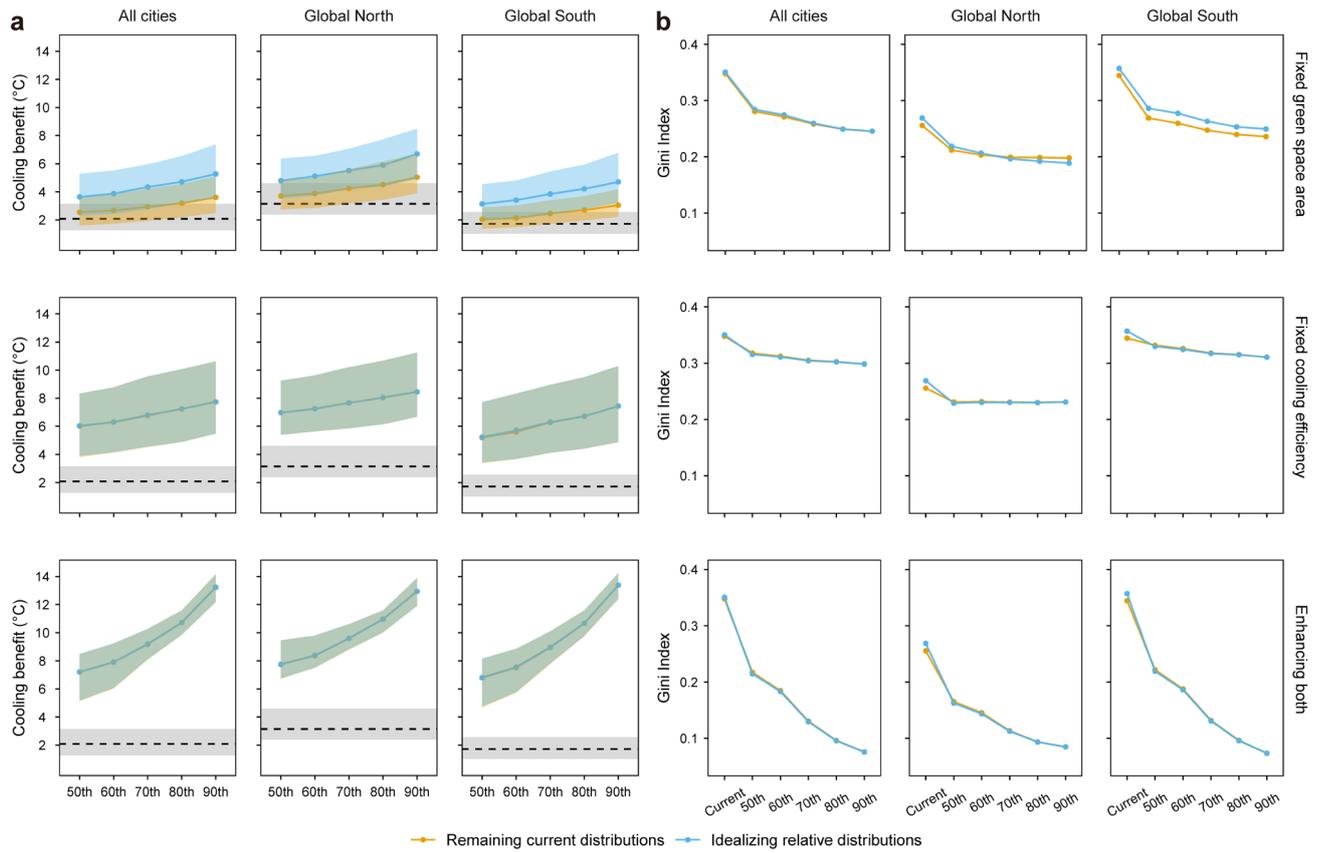

**Extended Data Fig. 10. Estimated potential of enhancing cooling benefit and reducing its inequality. *a*,** The potential of enhancing cooling capacity via either enhancing cooling efficiency (holding green space area fixed, upper panels), or increasing urban green space area (holding cooling efficiency fixed, middle panels) alone is much lower than that of enhancing both cooling efficiency and urban green space area (lower panels). The dashed horizontal lines and the gray bands represent median and 25-75[th] percentiles of cooling capacity of current cities. The colored curves (median) and bands (95% confidence interval) represent the effects of enhancing cooling efficiency, urban green space area, or both to 50-90[th] percentile of their assumed regional upper bounds, respectively. ***b*,** The potential of reducing cooling capacity inequality is also higher when enhancing both cooling efficiency and urban green space area. Gini index weighted by population density is used to measure inequality. Similar results were found for cooling capacity (Fig. 5). Cooling benefit is dependent on the relative distributions between green spaces and humans, we thus compared the effects of remaining current distributions ('Remaining current distributions',

yellow) and idealizing the relative distributions ('Idealizing relative distributions', blue). See Methods for details. Idealizing the relative distributions between green spaces and humans can enhance cooling benefit when enhancing cooling efficiency alone (holding green space area fixed, upper panels in *a*).

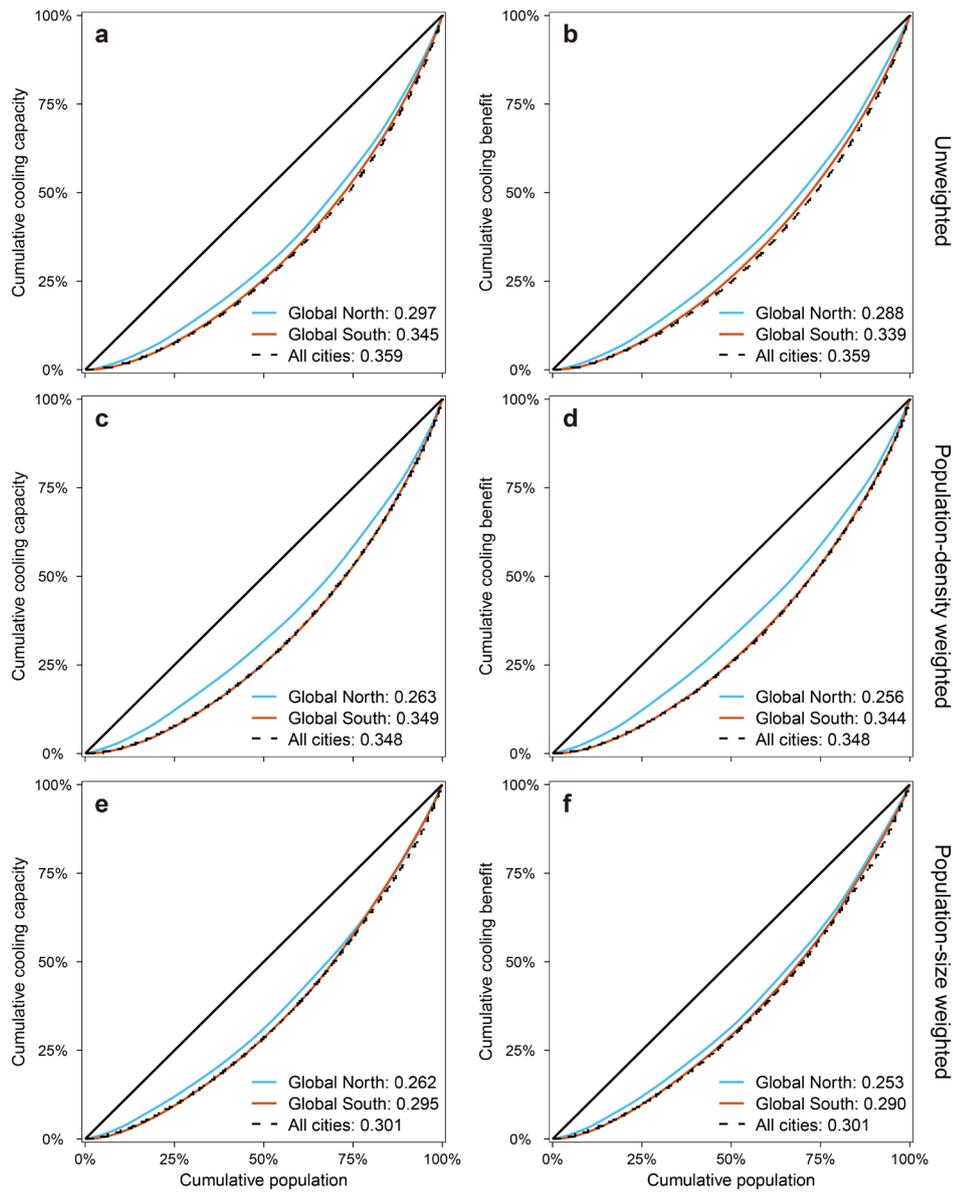

**Supplementary Fig. 1. The Lorenz curves for calculating the Gini coefficients of cooling capacity and cooling benefit.** *a, b,* Unweighted Gini. *c, d,* Gini weighted by population density. *e, f,* Gini weighted by population size. The Global South cities presented (population-density weighted) Gini coefficients of cooling capacity and cooling benefit (both around 0.35) that are ~30% higher than the Global North.

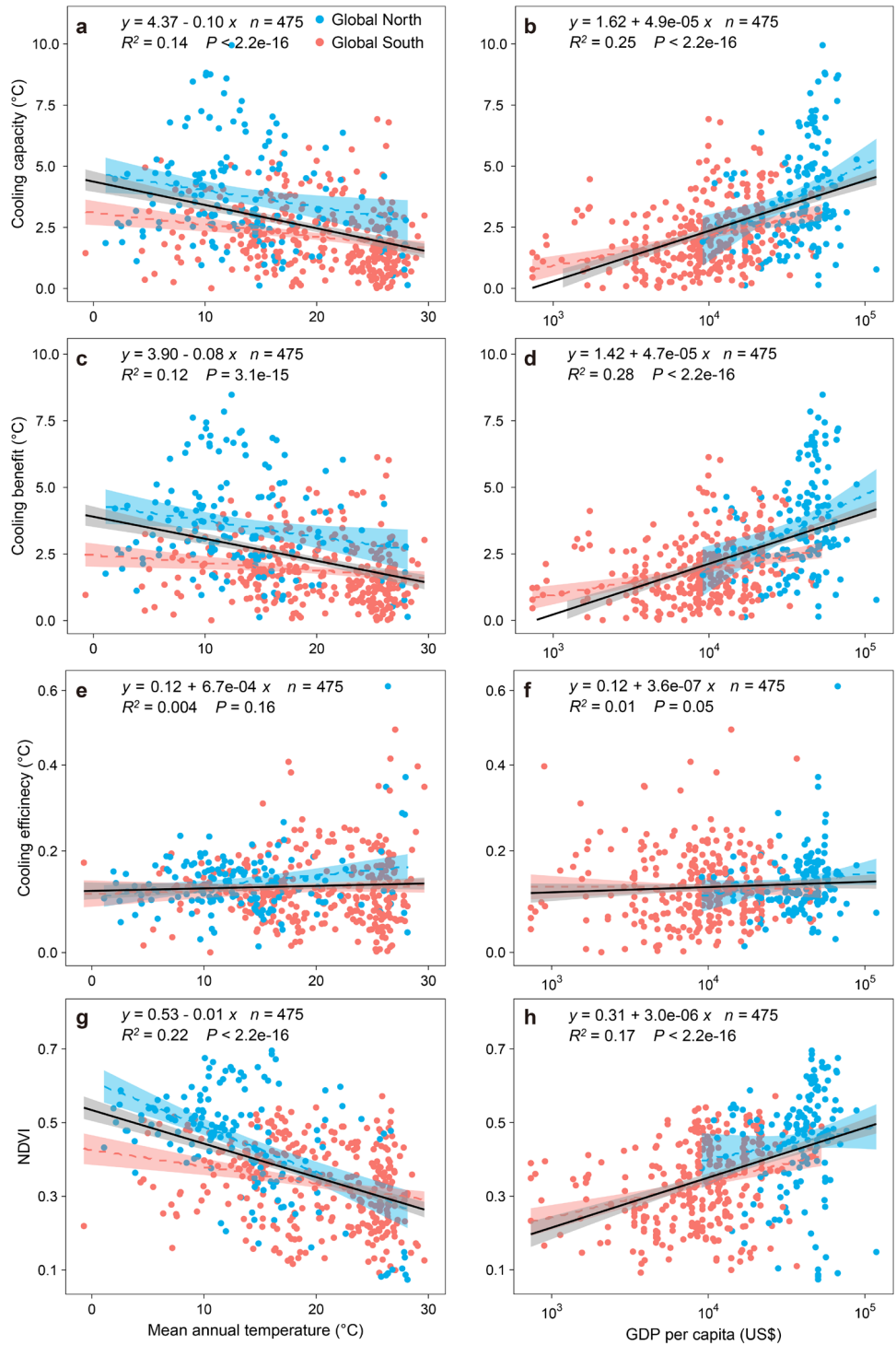

**Supplementary Fig. 2. Cooling capacity, cooling benefit, cooling efficiency and NDVI are significantly correlated with mean annual temperature (MAT) and GDP per capita.** The shaded areas represent 95% confidence intervals. We obtained $P$ values based on a two-sided test for the estimated slopes.

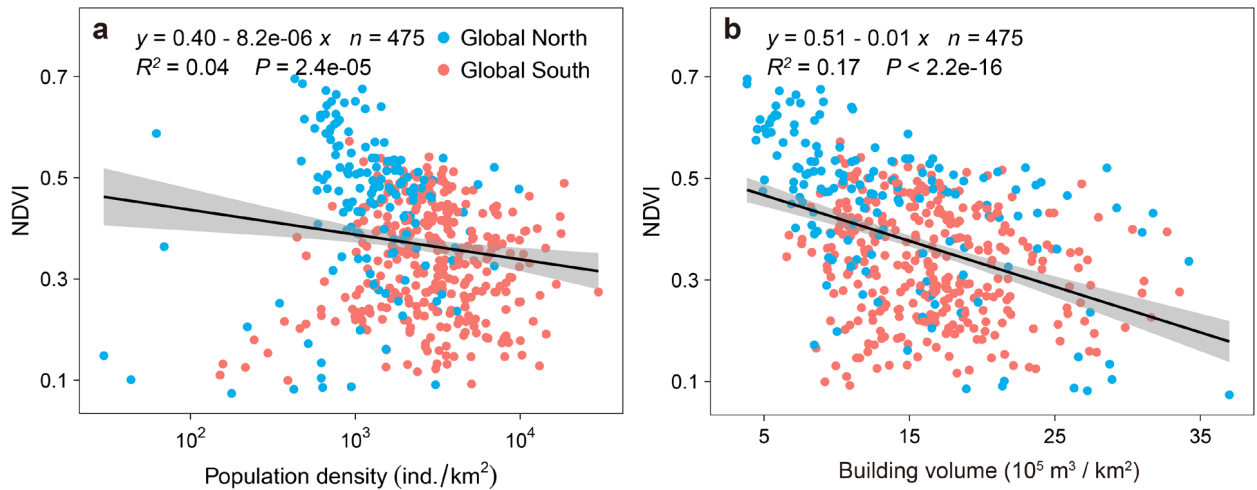

**Supplementary Fig. 3. NDVI is significantly correlated with population density and building volume.** *a,* Relationship between city-mean MODIS NDVI and population density. *b,* Relationship between city-mean MODIS NDVI and building volume. The shaded areas represent 95% confidence intervals. City-mean building volume is calculated using a global dataset on 3D built-up patterns of 2015 at 1 km resolution. We obtained *P* values based on a two-sided test for the estimated slopes.

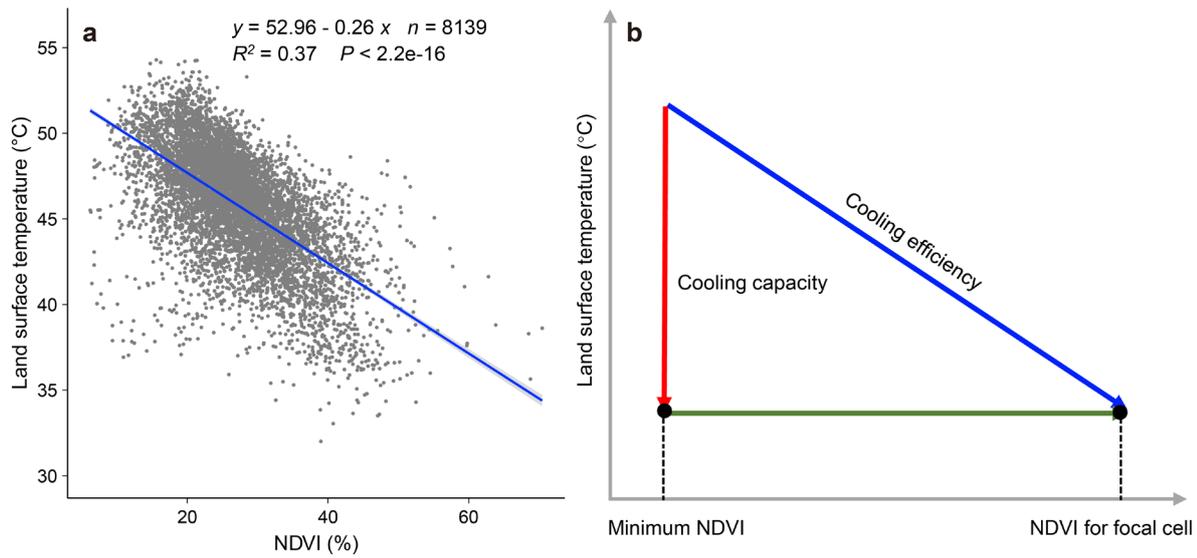

**Supplementary Fig. 4. Illustration of quantifying cooling efficiency and cooling capacity.** *a,* Relationship between land surface temperature and MODIS NDVI in the hottest month in 2018, taking Los Angeles as an example. Points represent 1-km MODIS pixels. The cooling efficiency slope of Los Angeles is 0.26 °C, meaning that per unit increase in green space corresponds to a reduction of 0.26 °C in land surface temperature. Linear regression line with 95% confidence interval is shown. *b,* Quantification of cooling capacity using NDVI and derived cooling efficiency in *a*. For a given grid cell, its local cooling capacity is calculated as the length of the red arrow, i.e., using minimum NDVI as the reference (see Supplementary Fig. 11 for using NDVI = 0 as the reference). The city-level cooling capacity is the mean value of local cooling capacity across the city, referred to as the average cooling effect provided by urban green spaces in comparison to the least vegetated areas. We obtained *P* values based on a two-sided test for the estimated slopes.

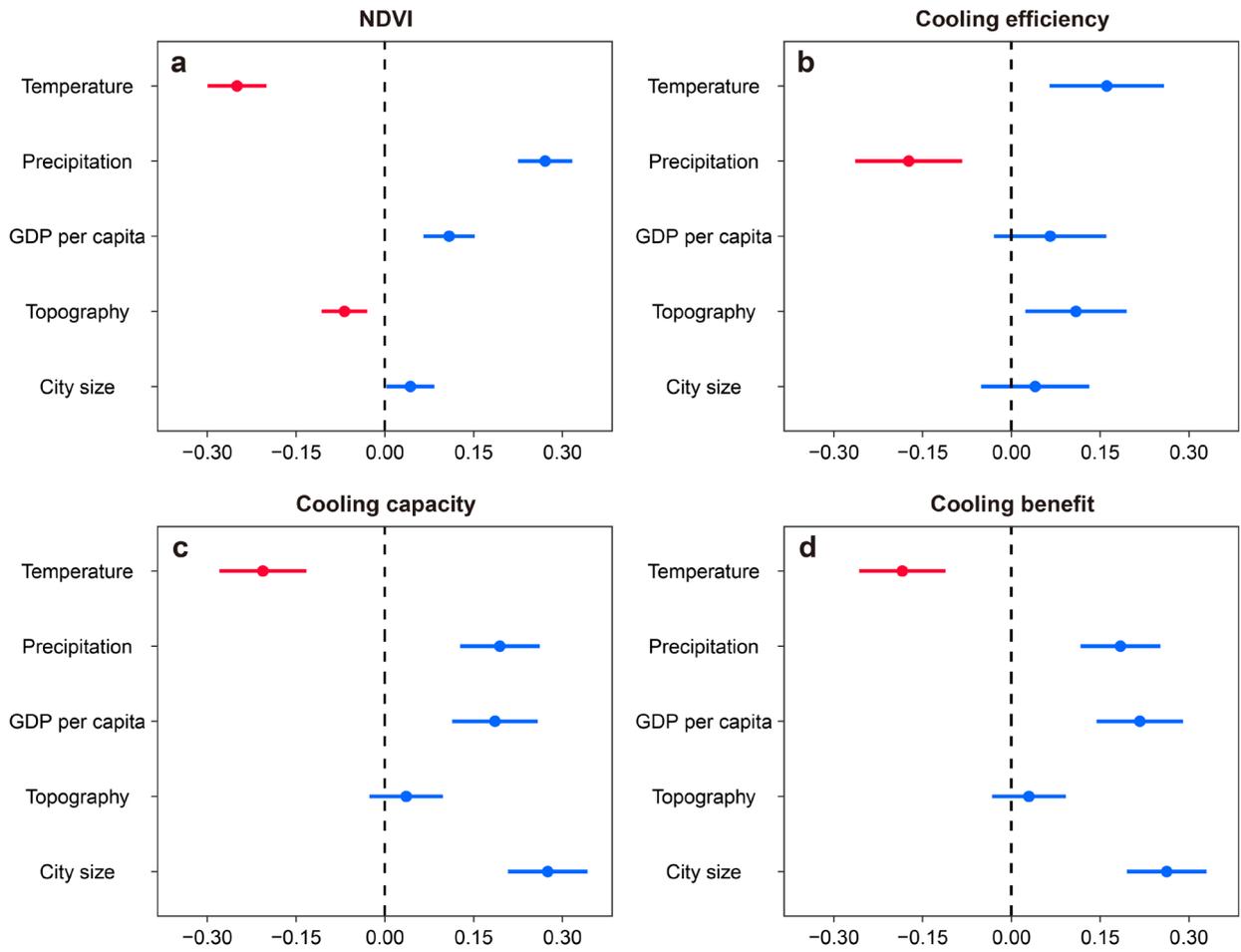

**Supplementary Fig. 5. Results from the Spatially Autoregressive Models (SAR) examining the effects of natural and socioeconomic factors. *a,*** NDVI. ***b,*** Cooling efficiency. ***c,*** Cooling capacity. ***d,*** Cooling benefit. We constructed SAR lag models with distance-based weight matrix as part of the spatial lag term. Consistent relationships were obtained with the non-spatial structural equation models (Fig 4, Extended Data Fig. 9).

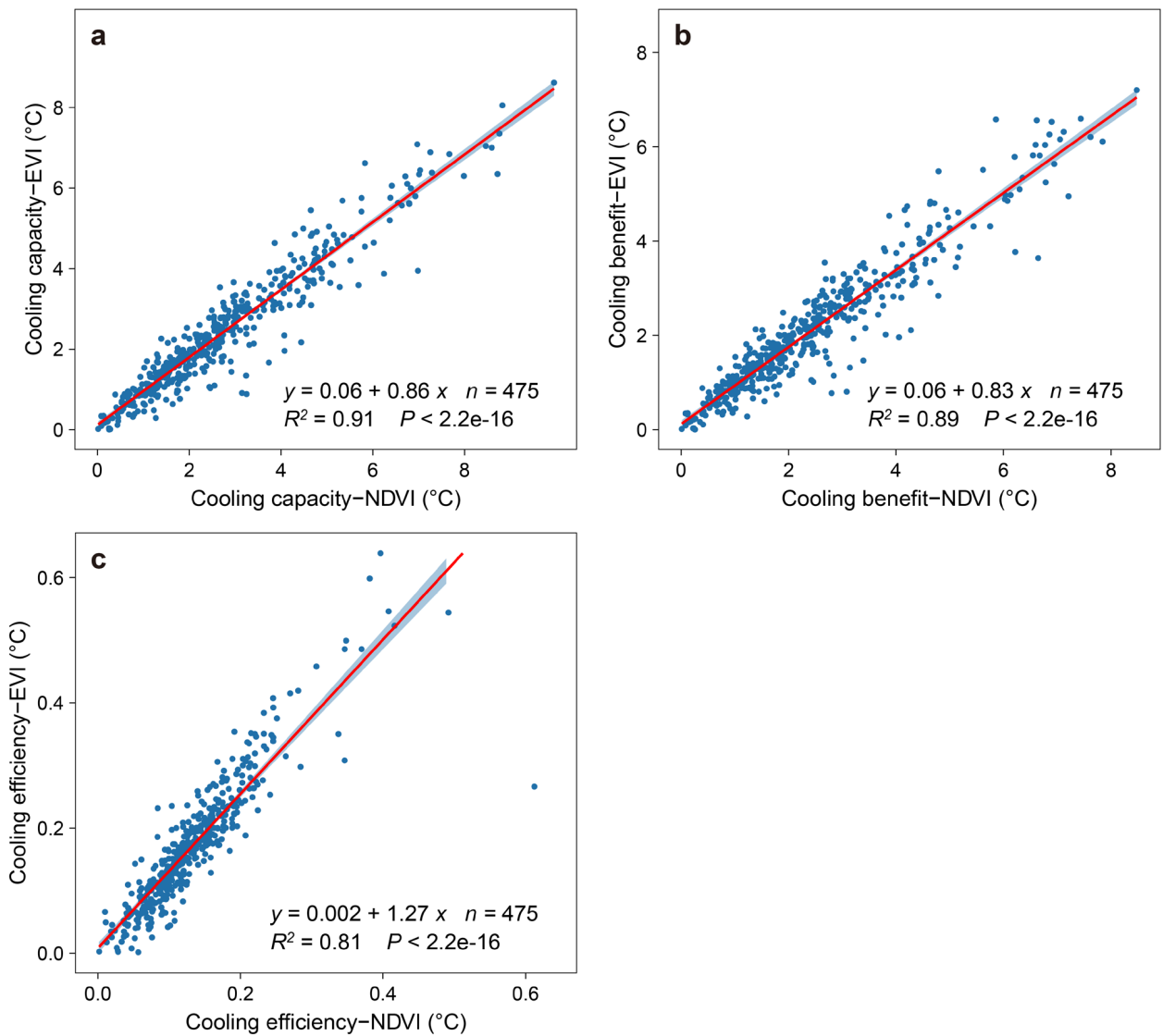

**Supplementary Fig. 6. Strong correlations between MODIS NDVI and MODIS EVI derived cooling capacity, cooling benefit and cooling efficiency.** *a,* Cooling capacity. *b,* Cooling benefit. *c,* Cooling efficiency. We obtained *P* values based on a two-sided test for the estimated slopes.

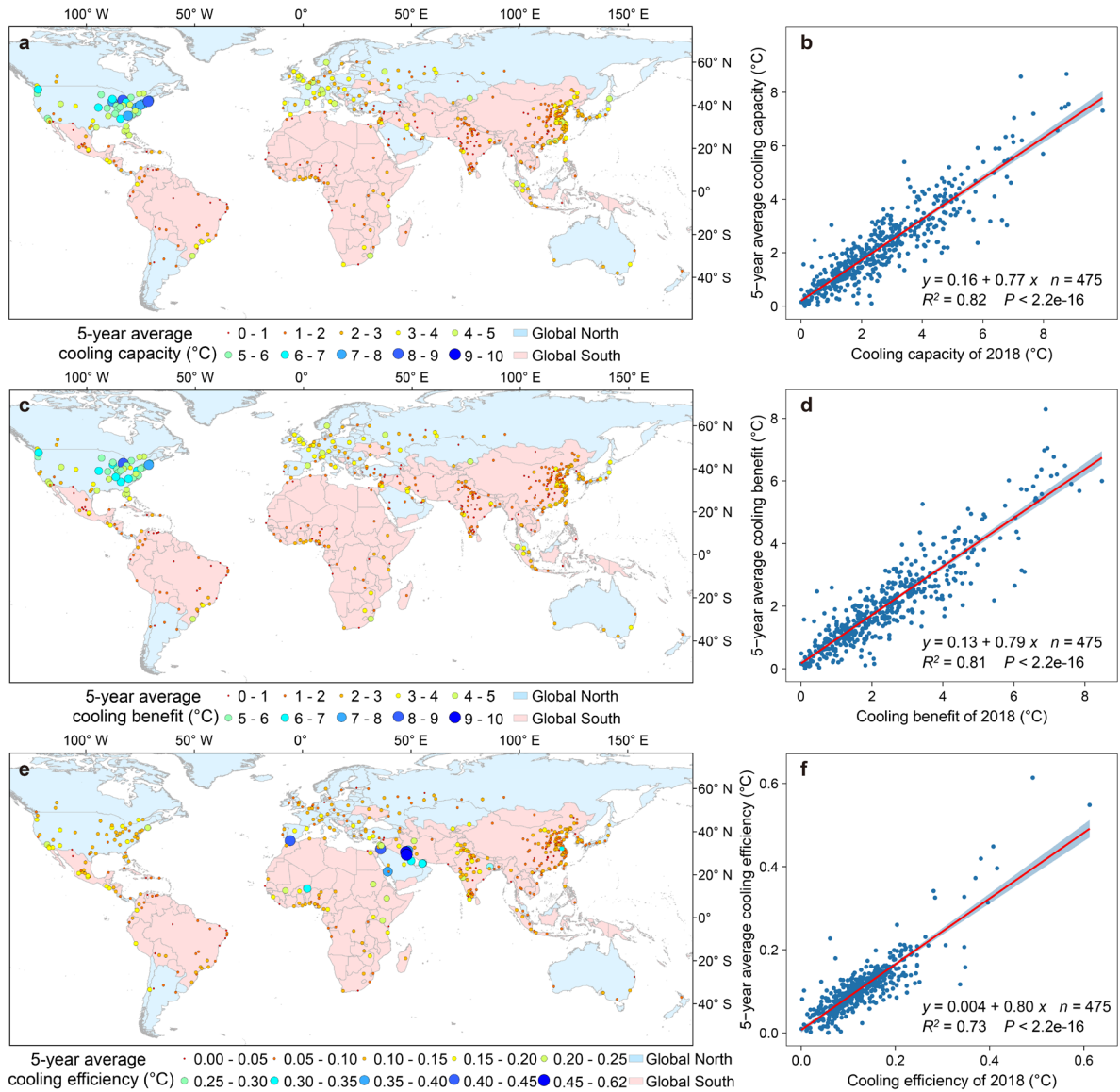

**Supplementary Fig. 7. Strong correlations between 1-year (2018) and 5-year (2014-2018) average cooling capacity, cooling benefit and cooling efficiency.** Global pattern of 5-year average cooling capacity (***a***), cooling benefit (***c***), and cooling efficiency (***e***) closely resembles that observed in the 1-year. We obtained *P* values based on a two-sided test for the estimated slopes.

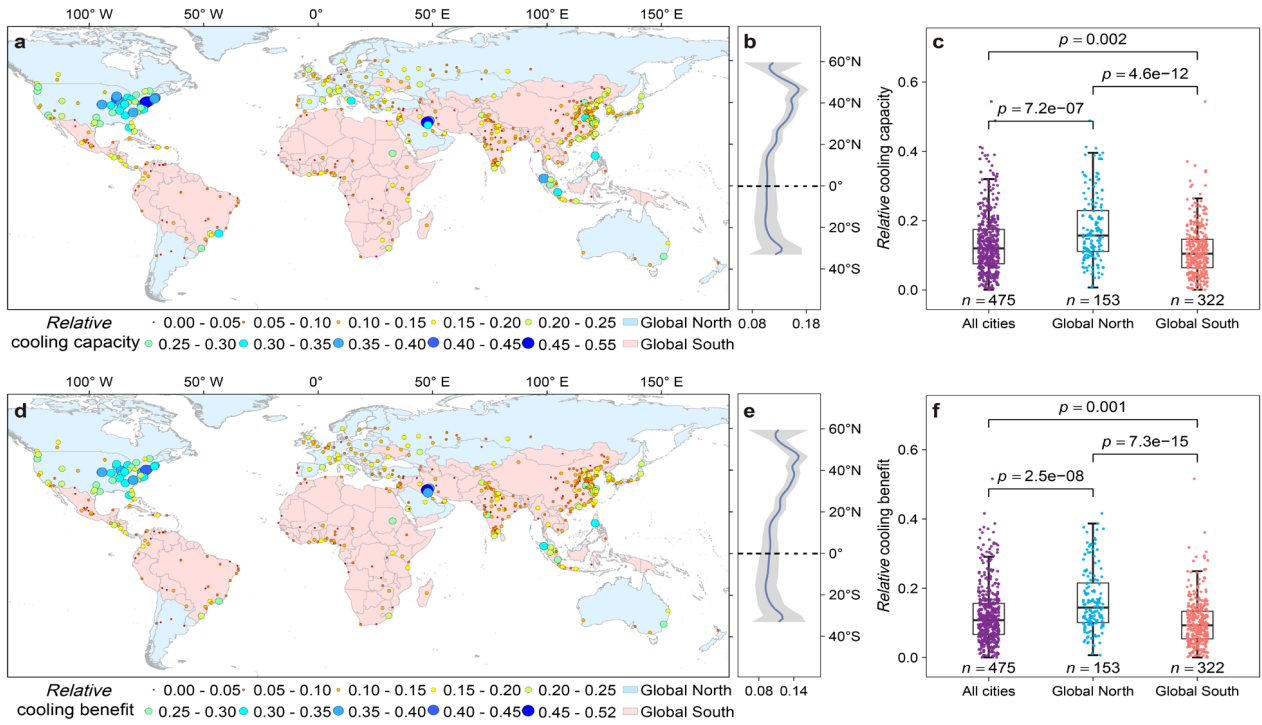

**Supplementary Fig. 8. Global pattern of *relative* cooling capacity and benefit.** *a-c,* Relative cooling capacity. *d-f,* Relative cooling benefit. This figure is generated using similar approach as in Fig. 1, but based on relative temperature cooled down (see Method). Cities with high cooling capacity/cooling benefit, such as those in the Northern America, also demonstrate a high relative cooling capacity and cooling benefit.

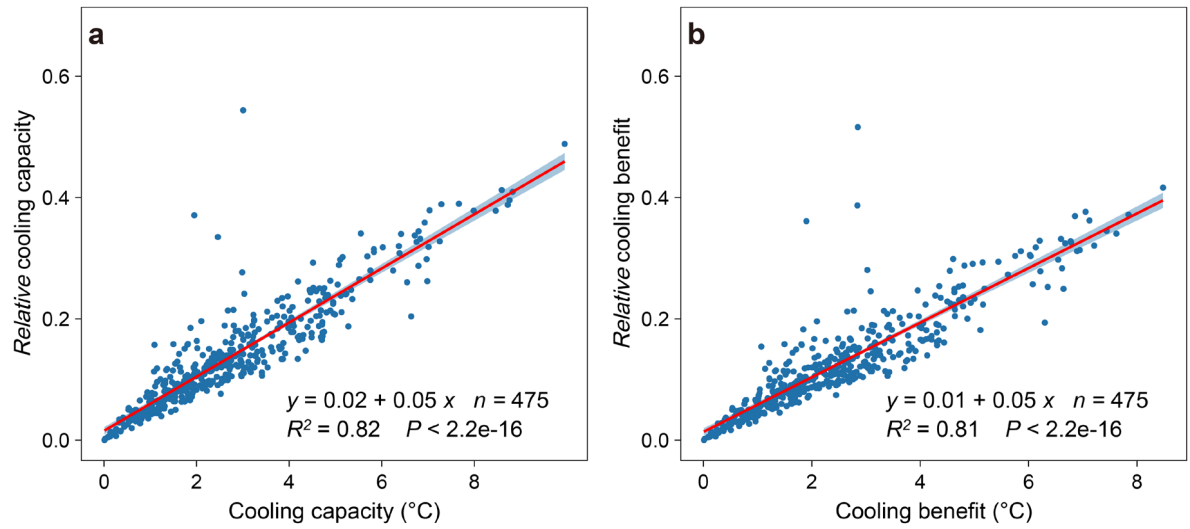

**Supplementary Fig. 9. Strong Correlations between absolute and *relative* cooling capacity/cooling benefit. *a,*** cooling capacity. ***b,*** cooling benefit. We obtained *P* values based on a two-sided test for the estimated slopes.

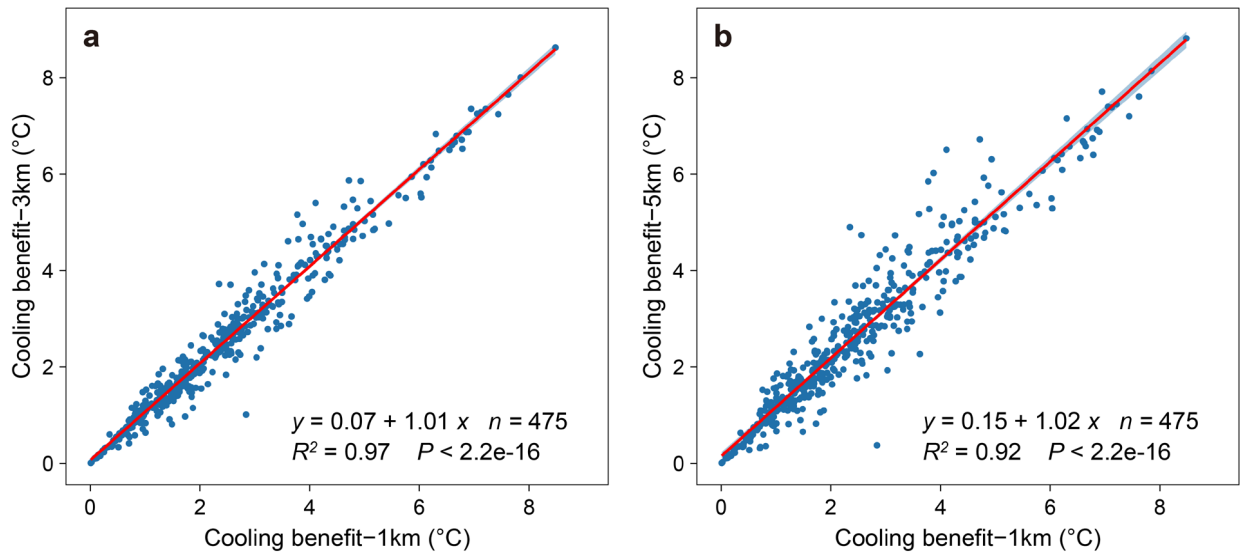

**Supplementary Fig. 10. Strong correlations between cooling benefit quantified at 1 km, 3 km and 5 km scales.** We obtained *P* values based on a two-sided test for the estimated slopes.

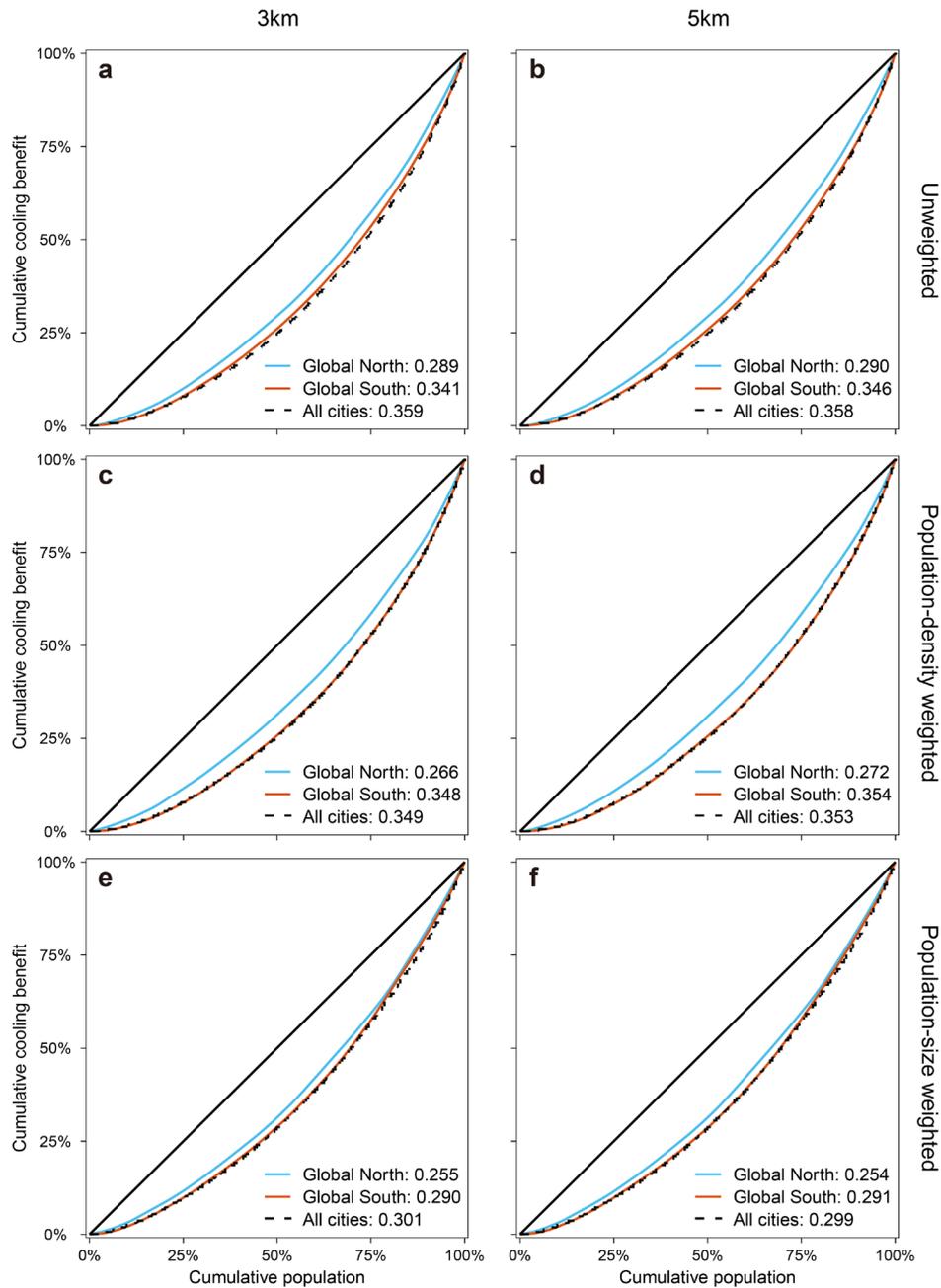

**Supplementary Fig. 11. The Lorenz curves and Gini coefficients of cooling benefit quantified at spatial scales of 3 km (left panel) and 5 km (right panel).** *a, b,* Unweighted Gini. *c, d,* Gini weighted by population density. *e, f,* Gini weighted by population size.

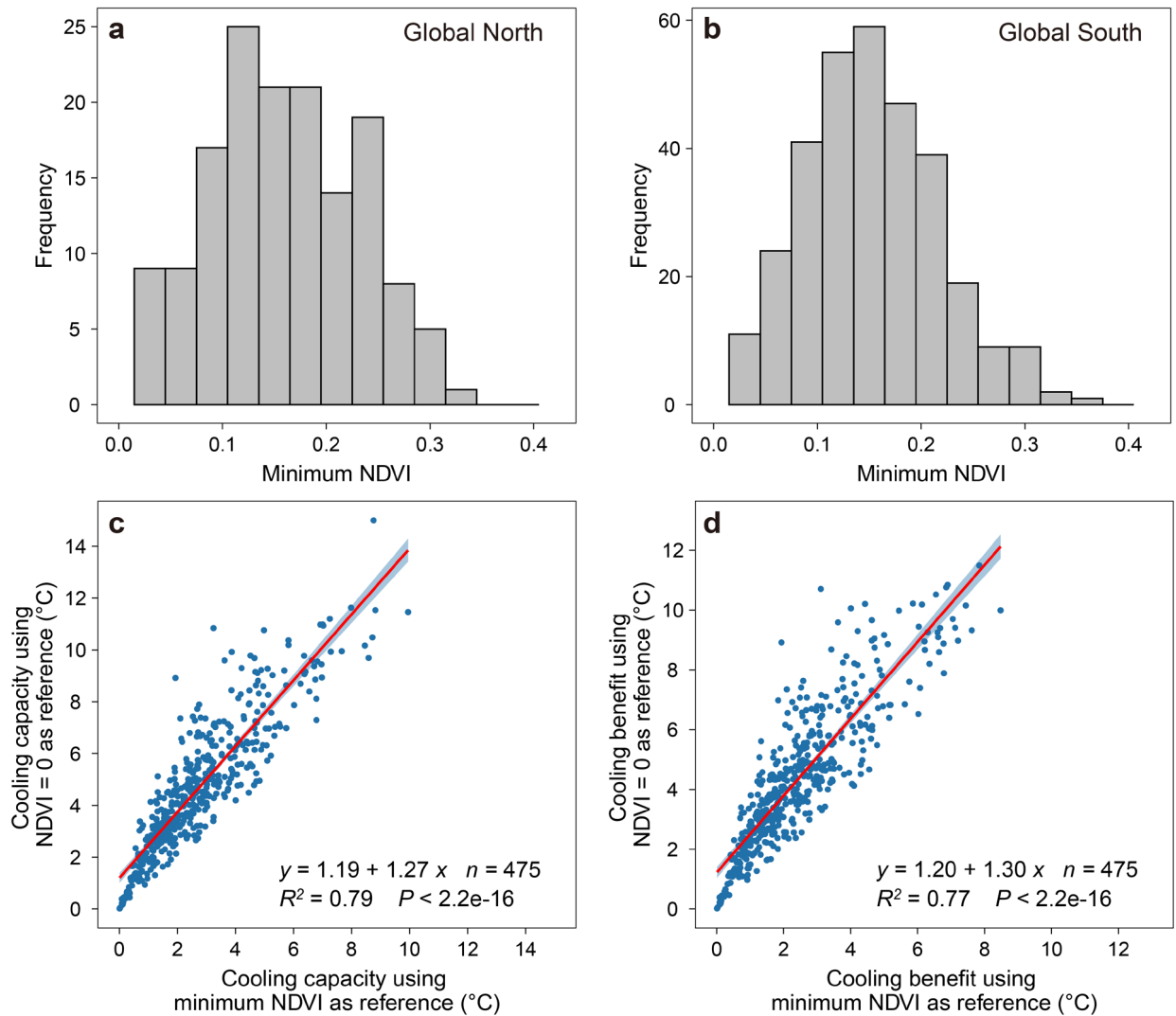

**Supplementary Fig. 12. Robustness analysis for using minimum NDVI values as references of cooling capacity quantification.** *a,* Histogram of minimum MODIS NDVI across the Global North and South cities. *b,* Histogram of minimum MODIS NDVI across the Global South cities. Most cities have minimum NDVI <0.25. *c,* Strong correlation of resulting cooling capacity between the methods using the minimum NDVI and NDVI = 0 as references. *d,* Strong correlation of resulting cooling benefit between the methods using the minimum NDVI and NDVI = 0 as references. See Supplementary Fig. 4 for method illustration. We obtained *P* values based on a two-sided test for the estimated slopes.

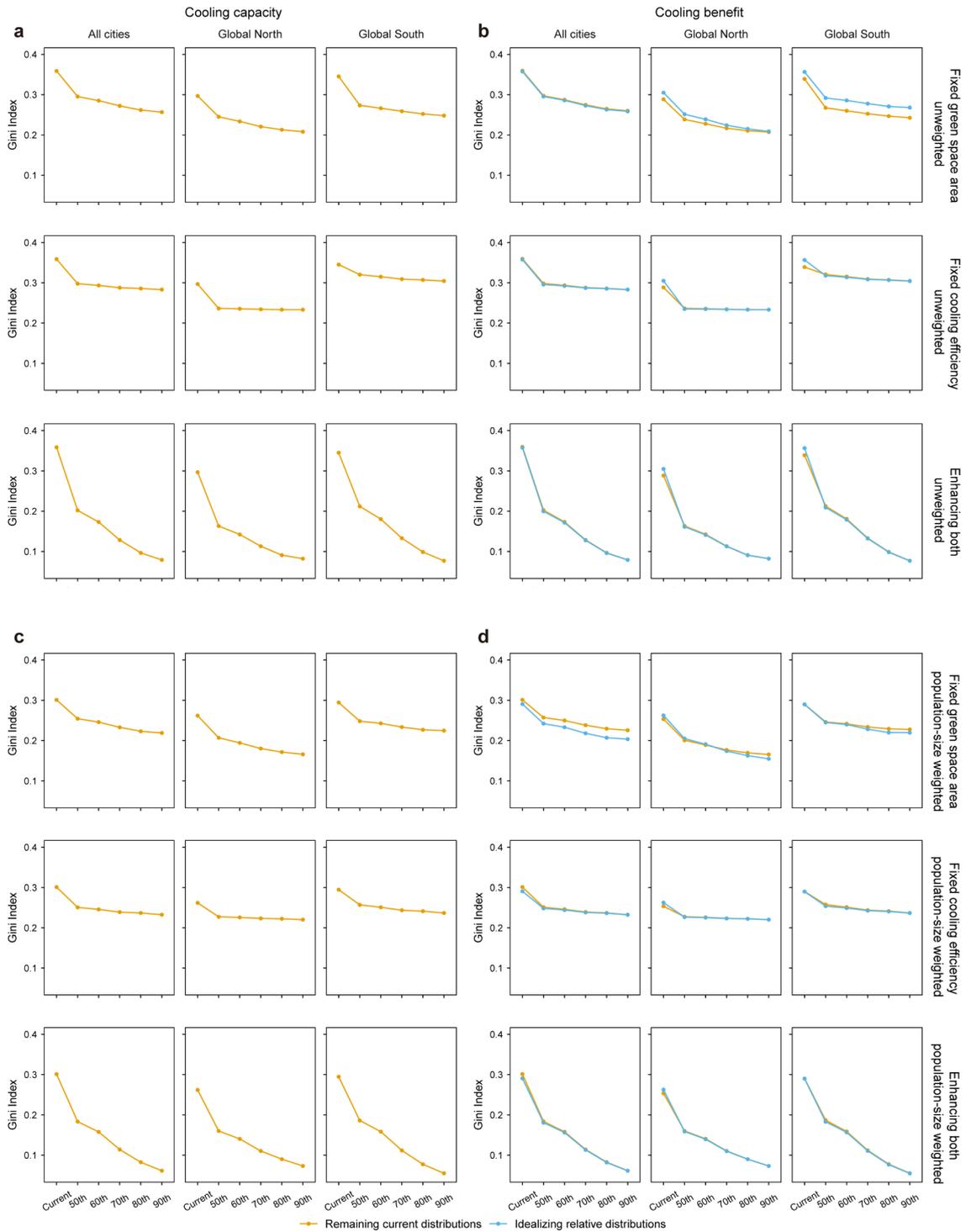

**Supplementary Fig. 13. Estimated potentials of reducing global inequality of cooling capacity and benefit.** This figure is generated using the same approach as in Fig. 5b but the Gini coefficients are calculated based on the unweighted and population-size weighted methods now. *a-b,* Unweighted Gini coefficients of potential cooling capacity and cooling benefit. *c-d,* Population-size weighted Gini coefficients of potential cooling capacity and cooling benefit.